\documentclass[amscd,amsmath,amssymb,verbatim, aps, prd, showpacs]{revtex4}
\usepackage{amsmath, amsthm, amssymb}

\usepackage{graphicx}
\usepackage[dvips]{hyperref}
\usepackage{textcomp}
\usepackage[OT1,T1]{fontenc}
\newtheorem{theorem}{Theorem}

\newtheorem{corollary}[theorem]{Corollary}

\begin{document}

\title{Generating Minimally Coupled Einstein-Scalar Field Solutions
from Vacuum Solutions with Arbitrary Cosmological Constant}
\author{Kjell Tangen}
\email{kjell.tangen@dnv.com}
\affiliation{DNV, 1322 H{\o}vik, Norway}\label{I1}
\date{\today}
\begin{abstract}
This paper generalizes two previously known techniques for generating
minimally coupled Einstein-scalar field solutions in 4 dimensions;
the Buchdahl and Fonarev transformations. Two generalizations are
made: i) the transformation is generalized to arbitrary dimension,
and ii) the new transformation allows vacuum solutions with non-zero
cosmological constant as seed. Thus, by applying this solution generation
technique, minimally coupled Einstein-scalar field solutions can
be generated from vacuum solutions with arbitrary cosmological constant
in arbitrary dimension. The only requirement to a seed solution
is that it posesses a hypersurface-orthogonal\ \ Killing vector
field.\ \ The generalization that allows us to use seed solutions
with arbitrary cosmological constant uncovers a new class of Einstein-scalar
field solutions that has previously not been studied. We apply the
new solution transformation to the $(A){\mathrm{dS}}_{4}$ vacuum
solution. Transforming the resulting Einstein-scalar field solution
to the conformal frame, a two-parameter family of spatially finite,
expanding and accelerating cosmological solutions are found that
are conformally isometric to the Einstein static universe $R\times
S^{3}$. We study null geodesics and find that for any observer,
the solution has a cosmological horizon at an angular distance of
$\pi /2$ away from the observer. A subset of these solutions are
studied in particular: A solution of this kind has an initial point
singularity that vanishes at early times as well as another point
singularity that emerges at late times. The solution is non-singular
in between those events. The late time singularity is hidden behind
an event horizon, and these solutions can therefore be naturally
interpreted as expanding cosmologies in which a scalar black hole
is formed at late times. The energy density is positive definite
only in parts of the parameter space. The conformally coupled scalar
field satisfies the weak energy condition as long as the energy
density is positive, while the strong energy condition is generally
violated.
\end{abstract}
\maketitle

\section{Introduction}\label{XRef-Section-51592622}

The physical relevance of scalar fields in today's gravitational
physics and cosmology primarily stems from i) their key role in
current models of early cosmological inflation, with predictions
that to an astonishing degree have been confirmed by recent cosmological
measurements \cite{Spergel:2006hy}, and ii) the viability of scalar
field models as candidate models for dark energy \cite{Ratra:1987rm}\cite{Caldwell:1997ii}\cite{Wang:1999fa,Ferreira:1997hj}.
The fact that scalar fields are inevitable artifacts of string theory
\cite{Polchinski-1998,Bertolami-Paramos-Turyshev-2006} provides
additional rationale for studying scalar fields. The continuing
focus on extra-dimensional models in fundamental physics provides
motivation for studying scalar fields in dimensions higher than
4.

There is an extensive literature on known Einstein-scalar field
solutions. In a recent paper, Wehus and Ravndal \cite{Wehus-Ravndal-2006}
provide a good historical overview of the various Einstein-scalar
field solutions relevant for the present paper. Solution generation
is a powerful method of discovering new solutions and uncovering
relationships between different solutions. There are numerous examples
of solutions to the Einstein-scalar field equations, which were
uncovered using solution generation techniques, that would be extremely
hard to derive by other means, including solving the field equations
by brute force. The way these solutions typiclly have been derived
is by first finding a new solution generation technique that transforms
a class of vacuum solutions into a class of Einstein-scalar field
solutions. Subsequently, this technique is applied to a known vacuum
solution of the Einstein equations to generate a new solution to
the Einstein-scalar field equations. Techniques for generating both
static as well as time dependent solutions are known by know.

Buchdahl \cite{Buchdahl:1959nk} derived static solutions to the
minimally coupled Einstein-scalar field equations in 4 dimensions
by applying a particular transformation to vacuum solutions of the
Einstein equations. He applied the solution generation technique
to the Schwarzschild solution and obtained the general static, spherically
symmetric solution to the minimally coupled Einstein-scalar field
equations for a massless scalar field. This particular solution
had been derived earlier by solving the field equations directly,
first by Bergmann and Leipnik \cite{Bergmann-Leipnik-1957} and later
by Janis, Newman and Winicour \cite{Janis-Newman-Winicour-1968}
.\ \ The Buchdahl transformation was later rediscovered by Janis,
Robinson and Winicour \cite{Janis-1969}.\ \ The static, spherically
symmetric scalar field solutions in 4 dimensions were generalized
to arbitrary dimension by Xanthopoulos and Zannias \cite{Xanthopoulos:1989kb}
by solving the Einstein-scalar field equations in arbitrary dimension.\ \

The action for a scalar field conformally coupled to Einstein gravity
in 4 dimensions was formulated by Callan, Coleman and Jackiw \cite{Callan-1970}.
Bocharova, Bronnikov and Melnikov found a static, spherically symmetric
black hole solution to the conformally coupled Einstein-scalar field
equations by directly solving the field equations. This solution,
however, was not known in the West until much later. Independently,
Bekenstein \cite{Bekenstein-1974} found a solution generation technique
that from an arbitrary solution to the massless, minimally coupled
scalar field equations in 4 dimensions could generate two solutions
to the massless, conformally coupled scalar field equations. He
first applied the Buchdahl transformation to the Scwharzschild solution
and then his solution generation technique to generate conformally
coupled, spherically symmetric solutions. The black hole solution
of Bocharova et. al. is among the generated solutions. This solution
is now known as the BBMB scalar black hole solution. The Bekenstein
solutions, including the BBMB black hole, were later rediscovered
by Fr{\o}yland \cite{Froyland:1982yd}.\ \ Maeda generalized the
Bekenstein solution technique to arbitrary dimension \cite{Maeda:1988ab}.
The Bekenstein technique is very general, and applies to any solution
to the minimally coupled scalar field equations.

Husain, Martinez and N\'unez found a time dependent, spherically
symmetric solution to the minimally coupled scalar field equations
with zero scalar potential \cite{Husain:1994uj}. This solution was
interpreted as a model of scalar field collapse. Soon after, Fonarev
found a technique for generating time dependent solutions to the
Einstein-scalar field equations in 4 dimensions from static vacuum
solutions to the Einstein equations \cite{Fonarev:1994xq} that extended
the Husain et. al solution to non-vanishing scalar potential. By
combining his technique with the Bekenstein transformation, he identified
a time dependent, conformally coupled solution that describes the
formation of a scalar black hole. Feinstein, Kunze and V\'azquez-Mozo
later generalized the Fonarev solutions to five dimensions \cite{Feinstein:2001xs}
.

In this paper, we will focus our attention on solutions to the minimally
coupled Einstein-scalar field equations in arbitrary dimension.
We will identify a class of solutions that includes both the Buchdahl
as well as the Fonarev solutions. This class of solutions can then
be combined with the Bekenstein transformation in order to generate
conformally coupled scalar field solutions.

In Section \ref{XRef-Section-5159214}, we review the minimally coupled
scalar field action and the Einstein-scalar field equations. Section
\ref{XRef-Section-421172432} reviews the general form of the vacuum
solutions we will use as seed for the transformation. In Section
\ref{XRef-Section-42223153}, we provide a brief review of the Buchdahl
transformation in 4 dimensions. Section \ref{XRef-Section-51592326}
introduces the classification scheme that later will be used to
label the different solutions. Section \ref{XRef-Section-51592441}\ \ contains
our main results, as it proves the new solution transformation.
The section ends by providing the new solution transformation in
a unified form that encapsulates both the Buchdahl and Fonarev transformations.
Finally, in Section \ref{XRef-Section-5159268}, we study an example
solution of the new kind, derived from an $(A){\mathrm{dS}}_{4}$
vacuum solution. There are two appendices: Appendix \ref{XRef-AppendixSection-98225849}
contains expressions that are useful for conformal transformations.
Appendix \ref{XRef-AppendixSection-510235039} provides, for refrence
purposes, the generalized Bekenstein theorem in arbitrary dimension
and for arbitrary scalar potential.\ \ \ \ \ \

\section{Einstein-Scalar Field Action and Field Equations}\label{XRef-Section-5159214}

This paper deals with solution generation techniques for minimally
coupled scalar fields. The solutions sought are solutions to the
minimally coupled coupled Einstein-scalar field equations. Let us
briefly review the Einstein-scalar field action and the field equations
derived from it when the action is\ \ extremalized. In $N$ dimensions,
the action for a scalar field minimally coupled to gravity is
\begin{widetext}\begin{equation}
S[ g,\phi ,\Lambda ,V] =\int d^{N}x\sqrt{-g}\left( \frac{1}{2\kappa
}\left( R[ g]  -2\Lambda \right) -\frac{1}{2}g^{\alpha \beta }\nabla
_{\alpha }\phi \nabla _{\beta }\phi -V[ \phi ] \right) ,
\end{equation}
\end{widetext}

where $\Lambda $ is a cosmological constant, $\kappa \equiv 8\pi
G$, $G$ being the $N$dimensional gravitational constant and $V[
\phi ] $ is an arbitrary scalar potential of the scalar field $\phi
$. Hereafter, we will include $\Lambda $ in the potential term,
defining an effective potential $W[ \phi ,\Lambda ] \equiv V[ \phi
] +\kappa ^{-1}\Lambda $. We get the minimally coupled Einstein-scalar
field equations by extremalizing the above action:
\begin{gather}
R_{\mu \nu }[ g] -\frac{1}{2}g_{\mu \nu }R[ g] =\kappa  \left( \nabla
_{\mu }\phi \nabla _{\nu }\phi -\frac{1}{2}g_{\mu \nu }g^{\alpha
\beta }\nabla _{\alpha }\phi \nabla _{\beta }\phi -g_{\mu \nu }W[
\phi ] \right) %
\label{XRef-Equation-9855237}
\\\Box \phi  -\frac{ d W[ \phi ] }{d \phi }=0.%
\label{XRef-Equation-9862429}
\end{gather}

$\nabla _{\alpha }$ is the covariant derivative associated with the
metric $g$, and $\Box  \equiv  \nabla ^{\alpha }\nabla _{\alpha }$
is the $N$ dimensional Laplace operator of this metric. $R_{\mu \nu
}[ g] $and $R[ g] $ are the Ricci tensor and Ricci scalar for an
arbitrary metric $g$.

By contracting equation \ref{XRef-Equation-9855237}, we can rewrite
this equation as
\begin{equation}
R_{\mu \nu }[ g] =\kappa ( \nabla _{\mu }\phi \nabla _{\nu }\phi
+\frac{2}{N-2}g_{\mu \nu }W[ \phi ] ) %
\label{XRef-Equation-9862442}
\end{equation}

\section{Orthogonal-Symmetric Geometries }\label{XRef-Section-421172432}

All solution transformations derived or referenced in this paper
use vacuum solutions as seed\footnote{A {\itshape seed solution}
is a known solution to the field equations that is used to generate
other solutions by use of a transformation, a {\itshape solution-generation
transformation, }which is a transformation that transforms one solution
into another solution, possibly to a different set of field equations.}.
By vacuum solution, we mean a solution to the vacuum Einstein equations
with an arbitrary cosmological constant $\Lambda $. We will require
the seed solutions to possess a hypersurface-orthogonal Killing
vector field \cite{Wald:1984-general}, in which case it is possible
to choose coordinates for which the line element associated with
the metric $g$ for the this geometry takes the following canonical
form:
\begin{equation}
{\mathrm{ds}}^{2}[ g] = \epsilon  e^{2U( x) }{\mathrm{dy}}^{2}+e^{-2U(
x) }h_{\mathrm{ij}}{\mathrm{dx}}^{i}{\mathrm{dx}}^{j}.%
\label{XRef-Equation-91972716}
\end{equation}

The coordinate $y$ is a parameter of the integral curves of the
Killing vector field, $\epsilon  = -1$ or $1$, depending on whether
the Killing vectors are time-like or space-like, $U( x) $ is a scalar
function that is independent of the $y$-coordinate and $e^{-2U(
x) }h$ is the metric of a foliation of $N-1$ dimensional hypersurfaces
orthogonal to the y-direction with coordinates $\{x_{i}, i=1,..,N-1\}$.
In the following, we will use $g[ U, h] $ to denote a metric written
on the form of eq. \ref{XRef-Equation-91972716}:
\begin{equation}
g_{00}[ U,h] = \epsilon  e^{2U( x) }, g_{\mathrm{ij}}[ U,h] =e^{-2U(
x) }h_{\mathrm{ij}}, g_{0i}[ U,h] =0%
\label{XRef-Equation-92192055}
\end{equation}

Throughout the paper, we will refer to this type of geometry as
a geometry with an {\itshape orthogonal symmetry} or being {\itshape
orthogonal-symmetric}. Likewise, we will refer to a metric of the
form of eq. \ref{XRef-Equation-92192055} as a metric with an {\itshape
orthogonal symmetry} or an {\itshape orthogonal-symmetric} metric.
Several well-known static vacuum solutions to the Einstein equations
are orthogonal-symmetric, including the Schwarzschild solution and
the Weyl solutions. The Kerr solution is an example of a space-time
geometry that is not orthogonal-symmetric. Although the Kerr solution
is stationary, which means that it has a time-like Killing vector
field, its $t=\mathrm{const}$ hypersufaces are not orthogonal to
the time-like Killing vector, and its metric can not be put on the
canonical form of eq. \ref{XRef-Equation-91972716}.

An important class of orthogonal-symmetric geometries is the class
of geometries with a time-like Killing vector orthogonal to a set
of space-like hypersurfaces. These are static geometries, which
posess metrics on the canonical form:
\[
{\mathrm{ds}}^{2}= - e^{2U( x) }{\mathrm{dt}}^{2}+e^{-2U( x) }h_{\mathrm{ij}}{\mathrm{dx}}^{i}{\mathrm{dx}}^{j}.
\]

Both the Buchdahl and Fonarev transformations were derived using
static seed solutions on this form \cite{Buchdahl:1959nk,Fonarev:1994xq}.

Because it will become useful in later derivations, let us evaluate
the Ricci tensor of an arbitrary orthogonal-symmetric metric written
on the form of equation \ref{XRef-Equation-91972716} in terms of
the scalar function $U( x) $, the $N-1$ dimensional metric $h$ and
its corresponding Ricci tensor $H_{\mathrm{ij}}\equiv R_{\mathrm{ij}}[
h] $. A lenghty calculation reveals the following expressions for
the Ricci tensor:
\begin{gather}
R_{00}[ g] =-\epsilon  e^{4U}( \nabla ^{2}U-\left( N-4\right) h^{\mathrm{kj}}\partial
_{k}U \partial _{j}U) %
\label{XRef-Equation-92082543}
\\R_{0i}[ g] =0%
\label{XRef-Equation-4247154}
\\R_{\mathrm{kj}}[ g] =H_{\mathrm{kj}}-\left( 6-N\right) \nabla
_{j}U \nabla _{k}U+h_{\mathrm{kj}}( \nabla ^{2}U-\left( N-4\right)
h^{\mathrm{il}}\partial _{i}U\ \ \partial _{l}U) +\left( N-4\right)
\nabla _{k}\nabla _{j}U,%
\label{XRef-Equation-9276244}
\end{gather}

where $\nabla ^{2}\equiv \nabla ^{i}\nabla _{i}\equiv
h^{\mathrm{kj}}\nabla _{k}\nabla _{j}$. Similarly, we may express
the $N$ dimensional Laplacian operator $\Box $\ \ in terms of the
$N-1$ dimensional Laplacian $\nabla ^{2}:$
\begin{equation}
\Box \phi =e^{2U( x) }( {\nabla }^{2}\phi -\left(
N-4\right) h^{\mathrm{il}}\partial _{i}U \partial _{l}\phi ) %
\label{XRef-Equation-9208260}
\end{equation}

\section{The Buchdahl transformation}\label{XRef-Section-42223153}

Before we treat the general case, let us briefly review the Buchdahl
transformation \cite{Buchdahl:1959nk}. Starting with a static vacuum
solution to the Einstein equations with vanishing cosmological constant
in 4 dimensions, the metric $\hat{g} = g[ V, h] $ can be written
on the form of eq. \ref{XRef-Equation-91972716}:\ \ \
\[
{\mathrm{ds}}^{2}[ \hat{g} ] = -e^{2V( x) }{\mathrm{dt}}^{2}+e^{-2V(
x) }h_{\mathrm{ij}}{\mathrm{dx}}^{i}{\mathrm{dx}}^{j},
\]

where $V( x) $ is a scalar function of the spatial coordinates $x^{i}$.
We will seek static solutions to the massless Einstein-scalar field
equations, eqs. \ref{XRef-Equation-9862429} and \ref{XRef-Equation-9862442}
with metrics on this form.\ \ Since $\hat{g}$ is the metric of a
vacuum solution to the Einstein equations, it follows from eq. \ref{XRef-Equation-92082543}\ \ that
$V$ must satisfy the 3 dimensional Laplace equation, $\nabla ^{2}V=0$.
Furthermore, by applying eq. \ref{XRef-Equation-9276244}, we find
that
\begin{equation}
H_{\mathrm{kj}}-2\nabla _{j}V \nabla _{k}V=0%
\label{XRef-Equation-42116542}
\end{equation}

Now, let us write the Einstein-scalar field equations eqs. \ref{XRef-Equation-9862429}\ \ and
\ref{XRef-Equation-9862442} for a general orthogonal-symmetric solution.
The scalar field equation, eq. \ref{XRef-Equation-9862429}, takes
the following form, using eq. \ref{XRef-Equation-9208260} and that
we are restricting our attention to a scalar field with zero potential:\ \ \ \
\begin{equation}
{\nabla }^{2}\phi =0
\end{equation}

The Einstein equations take the following form, using eqs. \ref{XRef-Equation-92082543}
and \ref{XRef-Equation-9276244} and assuming a static scalar field:
\begin{gather}
\nabla ^{2}U=0%
\label{XRef-Equation-42115556}
\\H_{\mathrm{kj}}-2\nabla _{j}U \nabla _{k}U+h_{\mathrm{kj}}\nabla
^{2}U=\kappa \nabla _{j}\phi \nabla _{k}\phi %
\label{XRef-Equation-421155516}
\end{gather}

Asserting that $\phi $ and $U$ both are functionals of $V$, i.e.
$\phi =F[ V] $ and $U=K[ V] $, we can rewrite eqs. \ref{XRef-Equation-42115556}
and \ref{XRef-Equation-421155516}:
\begin{gather}
\nabla ^{2}U=K^{{\prime\prime}}[ V] \nabla ^{\alpha }V\nabla _{\alpha
}V+K^{\prime }[ V] \nabla ^{2}V=0
\\H_{\mathrm{kj}}-\left( 2{K^{\prime }[ V] }^{2}+\kappa  {F^{\prime
}[ V] }^{2}\right) \nabla _{j}V\nabla _{k}V=0
\end{gather}

Using that that $V$ satisfies the Laplace equation $\nabla ^{2}V=0$
as well as eq. \ref{XRef-Equation-42116542}, we get the following
constraints for the functionals $F$ and $K$:\ \
\begin{gather*}
K^{{\prime\prime}}[ V] =0
\\\kappa  {F^{\prime }[ V] }^{2}+2{K^{\prime }[ V] }^{2}=2
\end{gather*}

This gives
\begin{gather*}
K=\alpha  V+K_{0}
\\F=\sigma  V+F_{0},
\end{gather*}

where $\alpha $, $K_{0}$ and $F_{0}$ are arbitrary constants and
$\sigma  $ is a constant that satisfies ${\sigma  }^{2}=\frac{2}{\kappa
}(1-\alpha ^{2})$. $K_{0}$ just gives a scale factor that can be
absorbed by rescaling the coordinates, so we may set $K_{0}=0$.
Assuming that $g$ is asymptotically flat, $F_{0}$ is the asymptotic
value of the field $\phi$, so we will set $F_{0}=0$ in order to
let the scalar field $\phi$ go to zero as $V\rightarrow 0$ at infinity.\ \ For
an arbitrary static vacuum solution $g[ V, h] $ to the 4 dimensional
Einstein equations,\ \ ($g[ U,h] $, $\phi$) is a solution to the
minimally coupled Einstein-scalar field equations with zero scalar
potential:
\begin{gather}
\phi =\sigma  V%
\label{XRef-Equation-421163941}
\\U=\alpha  V,%
\label{XRef-Equation-421163952}
\end{gather}

where $\alpha$ is an arbitrary dimensionless constant in the interval
$[-1,1]$, and $\sigma ^{2}\equiv \frac{2}{\kappa }(1-\alpha ^{2})$.
Eqs. \ref{XRef-Equation-421163941} and \ref{XRef-Equation-421163952}
define the Buchdahl soluions, and the Buchdahl transformation is
defined by the relationships between the scalar field $\phi$ and
metric functions $U$ and $V$ given by the functionals $F$ and $K$.
We see that the Buchdahl transformation is straight forward to prove
in 4 dimension. In other dimensions, this is, as we will see later,
no longer the case.

\section{Solution Classification Scheme}\label{XRef-Section-51592326}

In this section we will introduce the general classification scheme
that will be used throughout the paper to label different solutions
to the Einstein equations. The classification scheme is motivated
by the new solution transformation technique that will be derived
in the next section, and it labels scalar field solutions by characteristics
of the seed solution as well as properties of the transformation
applied to the seed solution. The new solution transformation is
a rather general result that encompasses two known solution transformation
techniques in 4 dimensions, the Buchdahl \cite{Buchdahl:1959nk}
and Fonarev\ \ \cite{Fonarev:1994xq} transformations.

The solution transformations referenced and derived in this paper
all share a common characteristic; they all derive minimally coupled
Einstein-scalar field solutions from orthogonal-symmetric vacuum
solutions to the Einstein equations, which according to the definition
of orthogonal-symmetric fields introduced in Section \ref{XRef-Section-421172432}
implies that the geometry of the seed solution must posesses a hypersurface-orthogonal
Killing vector field and, subsequently, the solution is invariant
with respect to translations along the integral curves of the Killing
vector field. The Buchdahl and Fonarev solution transformations
were derived for the special case of static seed solutions with
timelike, hypersurface-orthogonal Killing vector fields, but the
solution transformations extend easily to the more general case
of arbitrary hypersurface-orthogonal spacetime symmetries.

As seen in the previous section, the Buchdahl transformation \cite{Buchdahl:1959nk}
transforms an arbitrary 4-dimensional static vacuum solution $\hat{g}\equiv
g[ V,h] $ into a one-parameter family of static, minimally coupled
scalar field solutions with\ \ metrics $g=g[ U[ V] ,h] $. The metric
functions $U[ V] =\alpha  V$\ \ is parametrized by a a dimensionless
parameter $\alpha$ in the range $[-1,1]$.

The Fonarev transformation \cite{Fonarev:1994xq} extends the Buchdahl
transformation. It acts on the same class of seed vacuum solutions
as the Buchdahl transformation, which are solution metrics fitting
the metric template of eq. \ref{XRef-Equation-91972716}, but the
resulting scalar field solutions are time-dependent with metrics
that differ from eq. \ref{XRef-Equation-91972716} by a time-dependent
conformal factor:
\begin{equation}
{\mathrm{ds}}^{2}=e^{2\mu ( t) }( -e^{2U( x) }{\mathrm{dt}}^{2}+e^{-2U(
x) }h_{\mathrm{ij}}{\mathrm{dx}}^{i}{\mathrm{dx}}^{j}) %
\label{XRef-Equation-9266298}
\end{equation}

We will refer to metrics that fit this metric template as {\itshape
conformally inflated metrics}, or to be more specific, {\itshape
conformally time-inflated metrics, }referring to the time-dependent
conformal factor that transforms a static space-time geometry into
an inflating/deflating one. The geometry of eq. \ref{XRef-Equation-9266298}
is clearly conformally static. Let us generalize this geometry to
a geometry that is conformally ortogonal-symmetric:
\begin{equation}
{\mathrm{ds}}^{2}=e^{2\mu ( y) }( \epsilon  e^{2U( x) }{\mathrm{dy}}^{2}+e^{-2U(
x) }h_{\mathrm{ij}}{\mathrm{dx}}^{i}{\mathrm{dx}}^{j}) %
\label{XRef-Equation-42218146}
\end{equation}

The metric templates of eqs. \ref{XRef-Equation-91972716} and \ref{XRef-Equation-42218146}\ \ motivate
the introduction of a solution classification based on how the generated
solutions fit the template of eq. \ref{XRef-Equation-42218146}.\ \ Orthogonal-symmetric
solutions with $\mu =0$ represent one class of generated solutions,
hereafter given the label $\mathcal{S}$, whereas another class are
the $y$-dependent solutions with$ \mu \neq 0$, hereafter given the
label $\mathcal{T}$. Class $ \mathcal{T}$ solutions are hereafter
referred to as {\itshape conformally y-inflated} solutions, referring
to the y-dependent conformal factor $e^{2\mu ( y) }$ that represents
an inflation/deflation of a y-invariant geometry along the y-direction.

The Buchdahl and Fonarev transformations share a vital restriction:
Any seed solution for these transformations must be a vacuum solution
to the Einstein equations with vanishing cosmological constant ($\Lambda
=0$). If we are able to lift this restriction, i.e. derive scalar
field solutions from vacuum solutions with non-vanishing cosmological
constant, we get two new classes of solutions:
\begin{itemize}
\item $\mathcal{S}^{\Lambda }$: Orthogonal-symmetric solutions (with
$\mu =0$), derived from orthogonal-symmetric vacuum solutions with
$\Lambda  \neq  0$
\item $\mathcal{T}^{\Lambda }$: $y$-dependent, conformally y-inflated
solutions derived from orthogonal-symmetric vacuum solutions with
$\Lambda$ $ \neq $ 0
\end{itemize}

To be consistent with this labeling, we will label the solution
classes $ \mathcal{S}$ and $\mathcal{T}$ introduced above as $\mathcal{S}^{0}$
and $\mathcal{T}^{0}$, respectively, indicating that they derive
from seed solutions with $\Lambda =0$.

Let us summarize our scheme for labeling minimally coupled scalar
field solutions derived from vacuum solutions to the Einstein equations:
\begin{itemize}
\item $\mathcal{S}_{N}^{\Lambda }$: Orthogonal-symmetric solutions
($\mu =0$) derived from orthogonal-symmetric vacuum solutions with
$\Lambda$ $ \neq $ 0
\item $\mathcal{S}_{N}^{0}$: Orthogonal-symmetric solutions ($\mu
=0$) derived from orthogonal-symmetric vacuum solutions with $\Lambda$
= 0
\item $\mathcal{T}_{N}^{\Lambda }$: Conformally y-inflated solutions
($\mu  \neq  0$) derived from orthogonal-symmetric vacuum solutions
with $\Lambda$ $ \neq $ 0
\item $\mathcal{T}_{N}^{0}$: Conformally y-inflated solutions ($\mu
\neq  0$) derived form orthogonal-symmetric vacuum solutions with
$\Lambda$ = 0
\end{itemize}

\section{The New Solution Transformation}\label{XRef-Section-51592441}

\subsection{General solution constraints}

 From equation  \ref{XRef-Equation-9862442}\ \ we see that vacuum
solutions to the Einstein equations must satisfy the equation
\begin{equation}
R_{\mu \nu }[ g] =\frac{2}{N-2}g_{\mu \nu }\Lambda ,%
\label{XRef-Equation-92783053}
\end{equation}

with $\Lambda $ being an arbitrary cosmological constant. We will
take an arbitrary orthogonal-symmetric vacuum solution to the Einstein
equations, $\hat{g} = g[ V, \hat{h}] $, as our seed solution. Using
eq. \ref{XRef-Equation-91972716}, the metric $\hat{g}$ of this solution
can be written on the canonical form
\begin{equation}
d{\hat{s}}^{2}= \epsilon  e^{2V( x) }{\mathrm{dy}}^{2}+e^{-2V( x)
}{\hat{h}}_{\mathrm{ij}}{\mathrm{dx}}^{i}{\mathrm{dx}}^{j}.
\end{equation}

We are seeking to find the most general solution to the minimally
coupled Einstein-scalar field equations, eqs. \ref{XRef-Equation-9862429}
and\ \ \ref{XRef-Equation-9862442}, matching the metric template
of a conformally y-inflated metric:
\begin{equation}
{\mathrm{ds}}^{2}=e^{2\mu ( y) }(  \epsilon  e^{2U( x) }{\mathrm{dy}}^{2}+e^{-2U(
x) }h_{\mathrm{ij}}{\mathrm{dx}}^{i}{\mathrm{dx}}^{j}) , U( x) \equiv
\alpha  V( x) ,%
\label{XRef-Equation-926225147}
\end{equation}

where the $N-1$ dimensional metric $h_{\mathrm{ij}}$is conformally
related to ${\hat{h}}_{\mathrm{ij}}$ by a y-independent conformal
transformation:
\begin{equation}
{\hat{h}}_{\mathrm{ij}}=e^{2\omega ( x) }h_{\mathrm{ij}}, \omega
( x) \equiv \delta  V( x) ,%
\label{XRef-Equation-92785211}
\end{equation}

and posessing a scalar field on the form
\begin{equation}
\phi =\sigma  V( x) +\rho  \mu ( y) %
\label{XRef-Equation-92962746}
\end{equation}

with scalar field potential $W[ \phi ] $. For the time being, we
assume nothing about the potential, so it can be any function of
the scalar field.

In the above expressions, we have assumed that $W_{0}, \sigma ,
\rho , \alpha $ and $\delta $ are parameters that will be subject
to constraints in order to provide solutions to the field equations.

Before we proceed, let us relate the solution template defined by
equations \ref{XRef-Equation-926225147}-\ref{XRef-Equation-92962746}
to known solutions. For example, we can retrieve the Buchdahl solutions
\cite{Buchdahl:1959nk} in 4 dimensions, labeled $\mathcal{S}_{4}^{0}$,\ \ by
setting $\mu $ and $\Lambda $ to zero. Deriving our results in arbitrary
dimension, we should therefore immediately be able to generalize
the Buchdahl transformation from 4 to arbitrary dimension, i.e.
derive solutions of class $\mathcal{S}_{N}^{0}$.

Furthermore, if we set $\Lambda $ to zero, but allow $\mu $ to vary
with $y$, we retrieve the Fonarev class of solutions\cite{Fonarev:1994xq}
in 4 dimensions, labeled $\mathcal{T}_{4}^{0}$. Like for the Buchdahl
transformation, we should be able to generalize the Fonarev transformation
to arbitrary dimension, i.e. derive solutions of class $\mathcal{T}_{N}^{0}$.

Finally, the metric template of Equation \ref{XRef-Equation-926225147}
covers several classes of solutions that are not covered by previously
published results:
\begin{itemize}
\item $\mathcal{S}_{N}^{\Lambda }$: Orthogonal-symmetric solutions
derived from vacuum solutions with non-zero cosmological constant
\item $\mathcal{T}_{N}^{\Lambda }$: y-dependent, conformally inflated
solutions derived from vacuum solutions with non-zero cosmological
constant
\end{itemize}

Let $g\equiv e^{2\mu } \overline{g}$ denote the metric of equation
\ref{XRef-Equation-926225147} with $\overline{g}$ being an orthogonal-symmetric
metric; $\overline{g}\equiv g[ U, h] $. In the following, we will
use overdots to represent derivatives wrt. the coordinate y.

Applying the general conformal transformation formulas of Appendix
\ref{XRef-AppendixSection-98225849}, we get the following expressions
that relate the Ricci tensors of the metrics g and $\overline{g}$
:
\begin{gather}
R_{00} = {\overline{R}}_{00}-\left( N-1\right) \ddot{\mu }%
\label{XRef-Equation-9953847}
\\R_{0j}={}{}{\overline{{}R}}_{0j}+\left( N-2\right) \alpha \dot{\mu
}\partial _{j}V
\\R_{\mathrm{ij}}={\overline{{}R}}_{\mathrm{ij}}-{\overline{g}}_{\mathrm{ij}}
{\overline{g}}^{00}( \ddot{\mu }+\left( N-2\right) {\dot{\mu }}^{2})
\label{XRef-Equation-9953857}
\end{gather}

$\overline{\Box }\phi $ transforms as follows under a y-dependent
conformal transformation:
\begin{equation}
\Box \phi =e^{-2\mu }( \overline{\Box }\phi +\epsilon ( N-2)
e^{-2\alpha  V}\dot{\mu }\partial _{0}\phi ) =e^{-2\mu }( \sigma
\overline{\Box }V+\rho \overline{\Box }\mu
+\epsilon ( N-2)  e^{-2\alpha  V}\rho {\dot{\mu }}^{2}) %
\label{XRef-Equation-92763148}
\end{equation}

We will now proceed by first evaluating the Ricci tensor $R_{\mu \nu
}$using equations
\ref{XRef-Equation-9953847}-\ref{XRef-Equation-9953857}.\ \
${\overline{R}}_{\mu \nu }$can be derived from equations
\ref{XRef-Equation-92082543} - \ref{XRef-Equation-9276244}. When
evaluating $R_{\mu \nu }$, it is convenient to isolate terms
containing ${\hat{R}}_{\mu \nu }$ and use the fact that it is a
vacuum solution to the Einstein equations in order to reduce the
expressions as much as possible. Having evaluated the Ricci tensor,
we apply it to the Einstein equations, eq.
\ref{XRef-Equation-9862442}. In the course of development of the
equations, it is useful to expand the $N$-dimensional Ricci tensor
of an arbitrary orthogonal-symmetric metric $g=g[ U,h] $, $R_{\mu
\nu }[ g[ U,h] ] $,\ \ into expressions involving the metric
function $U$, the metric $N-1$ dimensional metric $h_{\mathrm{kj}}$
and its Ricci tensor $H_{\mathrm{ij}}\equiv R_{\mathrm{ij}}[ h] $.
Equations \ref{XRef-Equation-92082543} - \ref{XRef-Equation-9276244}
can be used to do that. The following expression for the Laplacian
$\Box \mu $ of an $y$-dependent function $\mu $ is also useful:\ \ \
\[
\Box \mu ( y) =\epsilon  e^{-2U}\ddot{\mu }
\]

After a lot of algebra, we find that the $(0,0)$ component of the
Einstein equations, eq. \ref{XRef-Equation-9862442}, reduces to
\begin{widetext}\begin{multline}
\left( N-1\right) \ddot{\mu }+\kappa  \rho ^{2} {\dot{\mu }}^{2}=\\
-\epsilon  {\alpha e}^{4\alpha  V}( 1-\alpha -\frac{\left( N-3\right)
}{N-4}\delta ) \left( N-4\right) h^{\mathrm{il}}\partial _{i}V \partial
_{l}V-\epsilon  \frac{2}{N-2} \left( \kappa  e^{2\mu }e^{2\alpha
V}W[ \phi ] - \alpha  e^{-2\left( 1-2\alpha -\delta \right) V}\Lambda
\right) %
\label{XRef-Equation-92963758}
\end{multline}
\end{widetext}

and the $(k,j)$ component of the Einstein equations reduces to
\begin{widetext}\begin{multline}
\left( \left( 6-N\right) +2\left( N-4\right) \delta -\left( N-3\right)
\delta ^{2}-\alpha ^{2}( 6-N) -\kappa  \sigma ^{2} \right) \partial
_{j}V \partial _{k}V\\
+\left( \left( N-3\right) \delta -\left( N-4\right) \left( 1-\alpha
\right) \right) \nabla _{k}\nabla _{j}V+\left( \alpha -\delta \right)
\left( \left( 1-\alpha \right) \left( N-4\right) -\left( N-3\right)
\delta \right)  h_{\mathrm{kj}}\nabla ^{l}V\nabla _{l}V=\\
\left( \frac{2\kappa }{N-2}e^{2\mu }e^{-2\alpha  V}W[ \phi ] -\frac{4-2\alpha
-2 \delta }{N-2}\Lambda \ \ e^{-2\left( 1-\delta \right) V} +\epsilon
\left( \ddot{\mu }+\left( N-2\right) {\dot{\mu }}^{2}\right)  e^{-4\alpha
V}\right) h_{\mathrm{kj}}%
\label{XRef-Equation-10495730}
\end{multline}
\end{widetext}

We see that in order for equation \ref{XRef-Equation-10495730} to
be satisfied with $\partial _{i}V \neq 0$, we must have
\begin{equation}
\delta =\frac{\left( N-4\right) }{\left( N-3\right) }\left( 1-\alpha
\right) %
\label{XRef-Equation-92964657}
\end{equation}

The (0,0) component of the Einstein equations then becomes
\begin{equation}
\left( N-1\right) \ddot{\mu }+\kappa  \rho ^{2} {\dot{\mu }}^{2}=-\epsilon
\frac{2}{N-2} \left( \kappa  e^{2\mu }e^{2\alpha  V}W[ \phi ] -
\alpha  \Lambda  e^{-2\left( 1-2\alpha -\delta \right) V}\right)
\label{XRef-Equation-424183122}
\end{equation}

Furthermore, since the left-hand side of equation \ref{XRef-Equation-92963758}
depends on $y$ only, it is evident that the only way to satisfy
this equation when $\dot{\mu }\neq 0$ and\ \ $\Lambda \neq 0$ is
to demand $1-2\alpha -\delta =0$. Since eq. \ref{XRef-Equation-92964657}
also must be satisfied, we find that when $\dot{\mu }\neq 0$ and\ \ $\Lambda
\neq 0$, $\alpha$ must take the unique value $\alpha =1/(N-2)$.

From eq. \ref{XRef-Equation-4247154} we find that $R_{0i}=0$ for
an orthogonal-symmetric metric. Therefore, the $(0,j)$ component
of eq. \ref{XRef-Equation-9862442} is identically satisfied if $\mu
=\mathrm{const}$. If $\mu \neq \mathrm{const}$, it gives us the
parametric constraint
\begin{equation}
\left( N-2\right) \alpha =\kappa  \rho  \sigma %
\label{XRef-Equation-106212126}
\end{equation}

Finally, the $(k,j)$ component of equation \ref{XRef-Equation-9862442}\ \ becomes,
having used equation \ref{XRef-Equation-92964657} to reduce it:
\begin{widetext}\begin{multline}
\left( \left( 6-N\right) +2\left( N-4\right) \delta -\left( N-3\right)
\delta ^{2}-\alpha ^{2}( 6-N) -\kappa  \sigma ^{2} \right) \partial
_{j}V \partial _{k}V=\\
\left( \frac{2\kappa }{N-2}e^{2\mu }e^{-2\alpha  V}W[ \phi ] -\frac{4-2\alpha
-2 \delta }{N-2}\Lambda \ \ e^{-2\left( 1-\delta \right) V} +\epsilon
\left( \ddot{\mu }+\left( N-2\right) {\dot{\mu }}^{2}\right)  e^{-4\alpha
V}\right) h_{\mathrm{kj}}
\end{multline}
\end{widetext}

Each side of this equation must vanish identically, so we get another
parametric constaint
\begin{equation}
\left( 6-N\right) +2\left( N-4\right) \delta -\left( N-3\right)
\delta ^{2}-\alpha ^{2}( 6-N) -\kappa  \sigma ^{2}=0%
\label{XRef-Equation-1097454}
\end{equation}

and the reduced equation
\begin{equation}
\ddot{\mu }+\left( N-2\right) {\dot{\mu }}^{2}=-\epsilon \frac{2\kappa
}{N-2}e^{2\mu }e^{2\alpha  V}W[ \phi ] +\epsilon \frac{4-2\alpha
-2 \delta }{N-2}\Lambda \ \ e^{-2\left( 1-\delta -2\alpha \right)
V}%
\label{XRef-Equation-9297151}
\end{equation}

Solving equation \ref{XRef-Equation-1097454} for $\sigma$, using
equation \ref{XRef-Equation-92964657}, gives $\sigma$ in terms of
$\alpha$:
\[
 \sigma ^{2}=\frac{\left( N-2\right)  }{\kappa ( N-3) }\left( 1-\alpha
^{2}\right)
\]

This is the same relation we found for the Buchdahl transformation
in section \ref{XRef-Section-42223153} in 4 dimensions. Now, if we
turn our attention to the equation of motion for the scalar field,
equation \ref{XRef-Equation-9862429},\ \ we can use equation
\ref{XRef-Equation-92763148} to compute $\Box \phi $. The scalar
field equation, eq. \ref{XRef-Equation-9862429}, then takes the form
\begin{equation}
\ddot{\mu } +\left( N-2\right) {\dot{\mu }}^{2}=\epsilon  \frac{1}{\rho
} e^{2\mu }e^{2\alpha  V}W^{\prime }[ \phi ] +\epsilon \ \ \frac{2}{N-2}\ \ \frac{\sigma
}{\rho }\Lambda  e^{-2\left( 1-2\alpha -\delta \right) V}%
\label{XRef-Equation-929720}
\end{equation}

We immediately see that the the left-hand sides of equations \ref{XRef-Equation-9297151}
and \ref{XRef-Equation-929720} are identical, and that, by equating
the right-hand sides of these equations, we get an equation constraining
the scalar field potential:
\begin{equation}
W^{\prime }[ \phi ]  =-\frac{2\kappa  \rho }{N-2}W[ \phi ] +\frac{2
\rho }{N-2}\left( 2-\alpha - \delta -\frac{\sigma }{\rho }\right)
\Lambda \ \ e^{2\left( \alpha +\delta -1\right) \frac{1}{\sigma
}\phi }e^{-2\left( \left( \alpha +\delta -1\right) \frac{\rho }{\sigma
}+1\right) \mu }%
\label{XRef-Equation-10617136}
\end{equation}

One way in which to satisfy Equation \ref{XRef-Equation-10617136}
is to have the second term on the right-hand side of this equation
vanish, which may happen when the following equation is satisfied:
\begin{equation}
W^{\prime }[ \phi ] =-\frac{2\kappa  \rho }{N-2}W[ \phi ] %
\label{XRef-Equation-1037126}
\end{equation}

and either $\frac{\sigma }{\rho }=2-\alpha -\delta  $\ \ or $\Lambda
=0$ (assuming $\rho \neq 0$). We will refer to this as Case 1. Equation
\ref{XRef-Equation-1037126} can easily be integrated, and gives
the exponential scalar field potential
\begin{equation}
W[ \phi ] =W_{0} e^{-2k \phi },%
\label{XRef-Equation-106211723}
\end{equation}

where $k\equiv \frac{\kappa  \rho }{N-2}$ and $W_{0}$ is an arbitrary
constant. Notice that, although equation \ref{XRef-Equation-1037126}
admits an additional constant term in the potential, equations \ref{XRef-Equation-9297151}
and \ref{XRef-Equation-929720} will not be identical unless this
constant term vanishes. Another way in which Equation \ref{XRef-Equation-10617136}
can be satisfied is that the $\mu $-dependence in the last term
vanishes, either by simply having $\mu =0$ (case 2), or by setting
$(\alpha +\delta -1)\frac{\rho }{\sigma }+1=0$ (case 3). Evidently,
it is not sufficient to require that equations \ref{XRef-Equation-9297151}
and \ref{XRef-Equation-929720} are identical; the equations must
also admit solutions. This adds more parametric constraints, because
it means that, provided $\partial _{i}V\neq 0,$ we must choose parameters
that makes the $V$-dependence of both eqs. \ref{XRef-Equation-424183122}
and \ref{XRef-Equation-9297151} vanish.

We will now show that equation \ref{XRef-Equation-106211723} is
the unique scalar field potential that satisfies equation \ref{XRef-Equation-10617136}
and is not in conflict with neither the constraints nor the $(0,0)$
field equation, eq. \ref{XRef-Equation-424183122}. This will prune
our solution space considerably, as we will have a unique expression
to use for the scalar field potential. The simplest case\ \ is case
2, i.e. $\mu =0$. In this case, we are able to derive the potential
directly from equation \ref{XRef-Equation-9297151}. Inserting this
expression in equation \ref{XRef-Equation-10617136} gives further
parametric constraints. When inserting the solutions into the (0,0)
field equation, eq. \ref{XRef-Equation-424183122}, we find that
the only solution admitted in this case is the potential of equation
\ref{XRef-Equation-106211723} with $\Lambda =0$. However, if both
$\Lambda =0$ and $ \mu =0$, eq. \ref{XRef-Equation-9297151} demands\ \ $W=0$.\ \ Finally,
considering case 3, we find the same thing. The potential must have
the exponential form of equation \ref{XRef-Equation-106211723},
and any case 3 solution with $\Lambda \neq 0$ is inconsistent with
the $(0,0)$ field equation.

Let us summarize what we have learnt so far:
\begin{theorem}

Given an orthogonally symmetric vacuum solution, $\hat{g}=$g[V,$\hat{h}$],
to the N dimensional Einstein equations wth an arbitrary cosmological
constant, equation \ref{XRef-Equation-92783053}, a scalar field
on the form \label{XRef-Theorem-103203143}
\begin{equation}
\phi =\sigma  V( x) +\rho  \mu ( y)
\end{equation}with a scalar field potential
\begin{equation}
W[ \phi ] =W_{0} e^{-2k \phi }
\end{equation}and a conformally y-inflated space-time with a metric
on the form
\begin{equation}
{\mathrm{ds}}^{2}=e^{2\mu ( y) }(  \epsilon  e^{2\alpha  V}{\mathrm{dy}}^{2}+e^{-2\alpha
V}e^{-2\delta  V}{\hat{h}}_{\mathrm{ij}}{\mathrm{dx}}^{i}{\mathrm{dx}}^{j})
\end{equation}is a solution to the minimally coupled Einstein-scalar
field equations, equations \ref{XRef-Equation-9862429} and \ref{XRef-Equation-9862442},
if and only if the following equations are satisfied:
\begin{gather}
\ddot{\mu }+\left( N-2\right) {\dot{\mu }}^{2}=-\epsilon \frac{2\kappa
W_{0}}{N-2}e^{2\left( 1-k \rho \right) \mu }e^{2\left( \alpha -k
\sigma \right) V}+\epsilon \frac{4-2\alpha -2 \delta }{N-2}\Lambda
\ \ e^{-2\left( 1-2\alpha -\delta \right) V}%
\label{XRef-Equation-103193029}
\\\ddot{\mu }+\frac{\kappa  \rho ^{2}}{\left( N-1\right) } {\dot{\mu
}}^{2}=-\epsilon  \frac{2}{\left( N-2\right) \left( N-1\right) }
\left( \kappa \ \ W_{0}e^{2\left( 1-k \rho \right) \mu }e^{2\left(
\alpha -k \sigma \right) V}- \alpha  \Lambda  e^{-2\left( 1-2\alpha
-\delta \right) V}\right) ,%
\label{XRef-Equation-103214451}
\end{gather}where $\Lambda$ is an arbitrary cosmological constant
and $\alpha$, $\rho$, $\sigma$, $\delta$, k, and $W_{0}$ are parameters
satisfying the following set of constraints:
\begin{gather}
\delta =\frac{\left( N-4\right) }{\left( N-3\right) }\left( 1-\alpha
\right) %
\label{XRef-Equation-10322143}
\\\sigma ^{2}=\frac{\left( N-2\right)  }{\kappa ( N-3) }\left( 1-\alpha
^{2}\right) .%
\label{XRef-Equation-10321522}
\end{gather}If $W_{0}=0$ and $\Lambda =0$, the parameter $\rho$
is given by
\begin{equation}
\rho ^{2}=\frac{1}{\kappa }\left( N-2\right) \left( N-1\right) %
\label{XRef-Equation-5264516}
\end{equation}If $W_{0}\neq 0$, the parameter k of the scalar field
potential is given in terms of $\rho$:
\begin{equation}
k=\frac{\kappa  \rho }{N-2}.%
\label{XRef-Equation-1098246}
\end{equation}If $\dot{\mu }$$ \neq $0, the following constraint
must be satisfied:
\begin{equation}
\left( N-2\right) \alpha =\kappa  \rho  \sigma  %
\label{XRef-Equation-103193044}
\end{equation}If $\Lambda $$ \neq $0, we must have
\begin{gather}
\frac{\sigma }{\rho }=2-\alpha -\delta %
\label{XRef-Equation-103214959}
\\1-2\alpha -\delta =0,%
\label{XRef-Equation-10982211}
\end{gather}which has the unique solution $\alpha =1/(N-2)$.
\end{theorem}

Equations \ref{XRef-Equation-103193029}-\ref{XRef-Equation-103193044}
do indeed cover a wide range of scalar field solutions, and we will
use the rest of this section to unravel the main solution classes.
As a result, we will be able to accurately label the Einstein-scalar
field solutions that can be generated by the solution transformation
of Theorem \ref{XRef-Theorem-103203143} in terms of the labeling
scheme introduced in the previous section.

Before we proceed, let us review the different cases to be considered:
\begin{itemize}
\item $\mu =0$ (solution class $\mathcal{S}_{N}$).
\begin{itemize}
\item $\Lambda$ $ \neq $0 (solution class $\mathcal{S}_{N}^{\Lambda
}$)
\item $\Lambda$=0 (solution class $\mathcal{S}_{N}^{0}$)
\end{itemize}
\item $\mu  \neq  \mathrm{const}$ (solution class $\mathcal{T}_{N}$)
\begin{itemize}
\item $\Lambda$ = 0 (solution class $\mathcal{T}_{N}^{0}$)
\item $\Lambda$$ \neq $0 (solution class $\mathcal{T}_{N}^{\Lambda
}$)
\end{itemize}
\end{itemize}

The various constraints set up by Theorem \ref{XRef-Theorem-103203143}
relate to the different solution classes as follows: All solutions
must satisfy equations \ref{XRef-Equation-10322143} and \ref{XRef-Equation-10321522}.
All class $\mathcal{T}$ solutions with non-vanishing scalar potential
must in addition satisfy equation \ref{XRef-Equation-103193044}.
$\mathcal{S}_{N}^{\Lambda }$and $\mathcal{T}_{N}^{\Lambda }$solutions
must satisfy equations \ref{XRef-Equation-103214959} and \ref{XRef-Equation-10982211},
while any solution with a non-vanishing scalar field potential must
satisfy equation \ref{XRef-Equation-1098246}.\ \

Let us consider for a moment the case of a seed vacuum solution
with constant metric function $V$. Without loss of generality, we
can set $V$ to zero. From equations \ref{XRef-Equation-92082543}
and \ref{XRef-Equation-92783053} we immediately see that the $(0,0)$
component of the Einstein equations, eq. \ref{XRef-Equation-9862442},
requires $\Lambda $ to be zero in this case, so this case falls
in the $\mathcal{S}_{N}^{0}$and $\mathcal{T}_{N}^{0}$solution classes.

\subsection{$S_{N}$: Orthogonal-symmetric scalar field solutions}

In the following two sections, we will state the individual sub
classes of $ \mathcal{S}$ and $ \mathcal{T}$ solutions in terms
of separate corollaries of Theorem \ref{XRef-Theorem-103203143}.

The $\mathcal{S}_{N}$classes of solutions emerge by setting $\mu
=0$ in Theorem \ref{XRef-Theorem-103203143}.\ \ In that case, equations
\ref{XRef-Equation-103193029} and \ref{XRef-Equation-103214451}
take the form
\begin{gather}
W_{0}e^{2\left( \alpha -k \sigma \right)  V}-\frac{\sigma }{\rho
} \frac{\Lambda }{\kappa } e^{2\left( 2\alpha  +\delta -1\right)
V}=0%
\label{XRef-Equation-10322628}
\\ W_{0}e^{2\left( \alpha -k \sigma \right)  V}- \alpha \frac{\Lambda
}{\kappa } e^{-2\left( 1-2\alpha -\delta \right) V}=0%
\label{XRef-Equation-10322635}
\end{gather}

Considering the $\mathcal{S}_{N}^{\Lambda }$class of solutions first,
i.e. assuming $\Lambda \neq 0$, we immediately find the constraint
$\frac{\sigma }{\rho }=\alpha $.\ \ In this case, equation \ref{XRef-Equation-103214959}\ \ implies
that $\alpha $ must be $1$. Setting $\alpha =1$ into equation \ref{XRef-Equation-10321522}
implies $\sigma =0$, in conflict with the constraint\ \ $\frac{\sigma
}{\rho }=\alpha $. Furthermore, we have to set $\Lambda =0$ in eqs.
\ref{XRef-Equation-10322628} and \ref{XRef-Equation-10322635}, we
must also have $W_{0}=0$. Thus, we have proved the following:
\begin{corollary}

$(\mathcal{S}_{N}^{\Lambda }, \mathcal{S}_{N}^{0})$ There are no
scalar field solutions of class $S_{N}^{\Lambda }$ that can be derived
using the transformation of Theorem \ref{XRef-Theorem-103203143},
i.e. there are no orthogonal-symmetric scalar field solution that
can be derived from a vacuum solution with cosmological constant
$\Lambda$$ \neq $ 0 using the solution transformation of Theorem
\ref{XRef-Theorem-103203143}. Furthermore, $W_{0}$ must be zero
for the entire $\mathcal{S}_{N}$ class of solutions, i..e there
are no solutions of class $\mathcal{S}_{N}$ with a potential term.\label{XRef-Corollary-1018173233}
\end{corollary}

Now, let us turn our attention to the $\mathcal{S}_{N}^{0}$solution
class, i.e. orthogonal-symmetric scalar field solutions with vanishing
cosmological constant. These are the generalized Buchdahl solutions
in dimension $N$. It is evident from eqs. \ref{XRef-Equation-10322628}
and \ref{XRef-Equation-10322635} that $W_{0}$ must also be zero,
i.e we are considering the case of a massless scalar field. The
remaining constraints in this case are equations \ref{XRef-Equation-10322143}
and \ref{XRef-Equation-10321522}, with the solution
\begin{equation}
\sigma =\pm \sqrt{\frac{\left( N-2\right) }{\kappa ( N-3)  } \left(
1-\alpha ^{2}\right) }=\frac{\beta }{\zeta },%
\label{XRef-Equation-103221949}
\end{equation}

where $ \zeta $ and $\beta $ are constants given by $\zeta ^{2}=
\kappa  \xi ^{2}$,$ \xi ^{2}=\frac{(N-2)}{4(N-1)}$ and $\beta ^{2}=\xi
^{2}( \frac{N-2}{N-3}) (1-\alpha ^{2})$. We can immediately see
that eq. \ref{XRef-Equation-103221949} reduces to Buchdahl's result
of eq. \ref{XRef-Equation-421163941} for $N=4$. Applying equation
\ref{XRef-Equation-103221949} to Theorem \ref{XRef-Theorem-103203143}
allows us to derive the characteristics of the\ \ $S_{N}^{0}$solution
class:
\begin{corollary}

$(\mathcal{S}_{N}^{0})$ If $\hat{g}=$g[V,$\hat{h}$] is an orthogonally
symmetric vacuum solution to the Einstein equations with vanishing
cosmological constant, we have a one-parameter family of solutions
to the minimally coupled Einstein-scalar field equations for a massless
scalar field given by an orthogonal-symmetric space-time metric
\label{XRef-Corollary-104611}
\begin{equation}
{\mathrm{ds}}^{2}= \epsilon  e^{2\alpha  V}{\mathrm{dy}}^{2}+e^{-2\gamma
V}{\hat{h}}_{\mathrm{ij}}{\mathrm{dx}}^{i}{\mathrm{dx}}^{j}
\end{equation}with a massless scalar field
\begin{equation}
\phi =\sigma  V,
\end{equation}where $\alpha$ is a free, dimensionless parameter
in the range [-1,1], $\gamma$$ \equiv $$\alpha +\frac{N-4}{N-3}(1-\alpha
)$ and $\sigma$$ \equiv $$\pm \sqrt{\frac{(N-2)}{\kappa ( N-3)
} (1-\alpha ^{2})}$.
\end{corollary}

Corollary \ref{XRef-Corollary-104611} was first proved in 4 dimensions
for stationary metrics by Buchdahl \cite{Buchdahl:1959nk} and later
by Janis, Robinson and Winicour \cite{Janis-1969}.\ \ We refer to
this version of the general transformation of Theorem \ref{XRef-Theorem-103203143}
as the {\itshape generalized Buchdahl transformation}.\ \ In 4 dimensions,
the Buchdahl transformation is easy to prove. However, to our knowledge,
this is the first time it has been generalized to arbitrary dimension.
Notice that for $N\neq 4$, the generalized Buchdahl transformation
of Corollary \ref{XRef-Corollary-104611} must do an extra $N-1$
dimensional conformal transformation $e^{2\delta  V}$of the $N-1$
dimensional hypersurface in order to provide a valid solution. This
implies that the $\mathcal{S}_{N}^{0}$scalar field solutions in
other dimensions than 4 become, in general, significantly more compelex
than the 4 dimensional solutions do.

\subsection{$\mathcal{T}_{N}:$ Conformally y-inflated scalar field
solutions}

Next, we will turn our attention to the $y$-dependent solutions,
i.e. the $\mathcal{T}_{N}$ solution class, and we will try to generate
$y$-dependent scalar field solutions from the constraints set up
by Theorem \ref{XRef-Theorem-103203143}. As noted above, all class
$ \mathcal{T}$ solutions must satisfy the constraints set up by
equations \ref{XRef-Equation-10322143} and \ref{XRef-Equation-10321522}.
$\mathcal{T}_{N}$ solutions with non-vanishing scalar potential,
i.e. $W_{0}\neq 0$, must in addition satisfy eq. \ref{XRef-Equation-103193044}.

\subsubsection{The metric function $\mu$(y)}

The y-dependence of the class $ \mathcal{T}$ solutions is defined
by the metric function $\mu ( y) $. Before we proceed to look at
specific classes of solutions that satisfy Theorem \ref{XRef-Theorem-103203143},
let us have a closer look at the differential equations that define
this function, eqs. \ref{XRef-Equation-10322628} and \ref{XRef-Equation-10322635}.
In the special case where both $W_{0}=0$ and $\Lambda =0$, these
equations both reduce, under the parametric constraint given by
eq. \ref{XRef-Equation-5264516}, to the equation
\begin{equation}
\ddot{\mu }+\left( N-2\right) {\dot{\mu }}^{2}=0,%
\label{XRef-Equation-53185222}
\end{equation}

which has the solution
\begin{equation}
\mu ( y) =\frac{1}{N-2}\ln  |\left( N-2\right) \sqrt{\eta }\left(
y-y_{0}\right) |%
\label{XRef-Equation-5464519}
\end{equation}

where $\eta $ and $y_{0}$\ \ are constants of integration \footnote{Notice
that, this particular form of the solution was chosen because, as
will become clear later, it fits a form that is common among several
of the class $ \mathcal{T}$ solutions.}.\ \ In all other cases,
eqs. \ref{XRef-Equation-10322628} and \ref{XRef-Equation-10322635}
are equivalent to the following two equations:
\begin{gather}
\ddot{\mu }- \left( 1-\lambda ^{2}\right) {\dot{ \mu }}^{2}= \left(
1-\lambda ^{2}\right) \chi %
\label{XRef-Equation-528652}
\\{\dot{\mu }}^{2}=\eta  e^{2\left( 1-\lambda ^{2}\right) \mu }-\chi
,%
\label{XRef-Equation-528726}
\end{gather}

where we have applied the constraints of Theorem \ref{XRef-Theorem-103203143}
and introduced the three parameters $\lambda $, $\eta $ and $\chi
$, defined as follows:
\begin{gather}
\lambda ^{2 }\equiv  k \rho =\left( N-3\right) \frac{\alpha ^{2}}{\left(
1-\alpha ^{2}\right) }%
\label{XRef-Equation-55142231}
\\\eta \equiv -\epsilon  \frac{2\kappa  W_{0}}{\left( N-2\right)
\left( N-1-\lambda ^{2}\right) }, \chi \equiv -\epsilon  \frac{2
\left( N-1-2 \alpha \right) }{\left( N-2\right) \left( N-3 \right)
\left( N-1-\lambda ^{2}\right) }\Lambda
\end{gather}

The presence of these three parameters reflects that, in general,
class $ \mathcal{T}$ solutions have three free parameters. One ($\lambda$,
or equivalently, $\alpha$) is shared with class $ \mathcal{S}$ solutions.
The second parameter, $\eta $, defines the zero-point value of the
effective scalar potential, $W[ \phi ] $. The last parameter, $\chi
$, is proportional to the cosmological constant, $\Lambda $. Equation
\ref{XRef-Equation-528652} follows by elimination of the potential
term, while eq. \ref{XRef-Equation-528726} follows by elimination
of the second-order derivative. Now, by taking the derivative of\ \ eq.
\ref{XRef-Equation-528726} and applying the value of $\alpha$ that
is valid for $\Lambda \neq 0$, $\alpha =1/(N-2)$, we get eq. \ref{XRef-Equation-528652}.
Thus, any solution to eq. \ref{XRef-Equation-528652} is also a solution
to eq. \ref{XRef-Equation-528726}.\ \ The general solution to these
equations is
\begin{equation}
\mu ( y) =\begin{cases}
\sqrt{-\chi }\left( y-y_{0}\right)  & \left( \eta =0, \chi <0\right)
\\
\sqrt{\eta -\chi }\left( y-y_{0}\right)  & \left( \eta >0, \eta
>\chi , \lambda ^{2}=1\right)  \\
\frac{1}{\left( \lambda ^{2}-1\right) }\ln |\left( 1-\lambda ^{2}\right)
\sqrt{\eta }\left( y-y_{0}\right) | & \left( \eta  >0, \chi =0\right)
\\
\frac{1}{\left( \lambda ^{2}-1\right) }\ln  |\sqrt{\frac{\eta }{\chi
}}\cos ( \left( 1-\lambda ^{2}\right) \sqrt{\chi }\left( y-y_{0}\right)
) | & \left( \frac{\eta }{\chi }>0, \chi \neq 0\right)  \\
\frac{1}{\left( \lambda ^{2}-1\right) }\ln  |\sqrt{\frac{\eta }{\chi
}}\sin ( \left( 1-\lambda ^{2}\right) \sqrt{\chi }\left( y-y_{0}\right)
) | & \left( \eta >0, \chi <0\right)  \\
\end{cases}%
\label{XRef-Equation-5465721}
\end{equation}

where $y_{0}$ is an arbitrary constant of integration. Notice that
there are no real solution for $\mu $ with parameter values $\eta
<0$ and $\chi  > 0$.

\subsubsection{Solution reparametrization}

The three parameters $\lambda $, $\eta $ and $\chi $ is a convenient
set of parameters. $\lambda $ is a dimensionless parameter that
takes any real value. Let us express the parameters used by Theorem
\ref{XRef-Theorem-103203143} in terms of the new parameters. Applying
the parametric constraints stated by Theorem \ref{XRef-Theorem-103203143},
the parameters $\alpha $ and $\sigma $ can conveniently be expressed
in terms of the dimensionless parameter $\lambda $:
\begin{gather}
\alpha =\varepsilon \frac{\lambda }{\sqrt{N-3+\lambda ^{2}}}, \sigma
= \varepsilon \sqrt{\frac{\left( N-2\right) }{\kappa ( N-3+\lambda
^{2}) }},%
\label{XRef-Equation-54222627}
\end{gather}

where $\varepsilon =\pm 1$. This parametrization of $\alpha $ and
$\sigma $ in terms of $\lambda $ is valid for all solutions of classes
$ \mathcal{S}$ and $ \mathcal{T}$. Similarly, $\rho $ and $k$ can
be expressed in terms of $\lambda $ as follows:
\begin{equation}
 \rho =\sqrt{\frac{\left( N-2\right) }{\kappa }}\lambda , k=\sqrt{\frac{\kappa
}{N-2}}\lambda %
\label{XRef-Equation-55142955}
\end{equation}

Finally, we can express the parameters $W_{0}$ and $\Lambda$ in
terms of the new parameters $\lambda $, $\eta $ and $\chi $:
\begin{equation}
W_{0}=-\epsilon  \frac{\left( N-2\right) }{2\kappa }\left( N-1-\lambda
^{2}\right)  \eta  ,\ \ \Lambda =-\epsilon \frac{ \left( N-2\right)
\left( N-3 \right) }{2 \left( N-1-2 \alpha \right) }\left( N-1-\lambda
^{2}\right) \chi %
\label{XRef-Equation-55165629}
\end{equation}

\subsubsection{Class $\mathcal{T}_{N}^{0}$ solutions with vanishing
scalar potential}

Let us first look at the $\mathcal{T}_{N}^{0}$class of solutions,
i.e. the solutions with $\Lambda =0$. There are two sub classes
of solutions that we must treat separately: $W_{0}=0$ and $W_{0}\neq
0$. The simplest case is $\mathcal{T}_{N}^{0}$ with vanishing scalar
potential; $W_{0}=0$. In this case, the right hand sides of the
$(0,0)$ and $(k,j)$ components of the Einstein equations, equations
\ref{XRef-Equation-103214451} and \ref{XRef-Equation-103193029},
both vanish. By comparing the left hand sides of these equations,
we see that the constraint $\kappa  \rho ^{2}=(N-1)(N-2)$ must be
fulfilled in order for these two equations to not be in conflict
and still yield solutions with $\dot{\mu }\neq 0.$ The field equations
reduce to the differential equation of eq. \ref{XRef-Equation-53185222},
which has the general solution $\mu ( y) =\frac{1}{N-2}\ln  |\sqrt{\eta
}(y-y_{0})|$. Transforming to a new coordinate $Y$ given by $|(N-2)\sqrt{\eta
}(y-y_{0})|^{\frac{2}{N-2}}=e^{2\sqrt{\eta }Y}$ allows us to reexpress
the solution on another form. Let us summarize this set of solutions:
\begin{corollary}

 $(\mathcal{T}_{N}^{0}, W_{0}=0)$ The $\mathcal{T}_{N}^{0}$ class
of solutions with vanishing scalar field potential consists of a
one-parameter family of solutions. The metric is given by \label{XRef-Corollary-101818159}
\begin{equation}
{\mathrm{ds}}^{2}=\epsilon  {e^{2\alpha  V+}}^{2\lambda ^{2}\sqrt{\eta
}Y}dY^{2}+e^{2\sqrt{\eta }Y}e^{-2\gamma  V}{\hat{h}}_{\mathrm{ij}}dx^{i}dx^{j},
\end{equation} where the parameters $\alpha $, $\sigma $ and $\rho
$ are given by eqs. \ref{XRef-Equation-54222627} and \ref{XRef-Equation-55142955}
with $\lambda =\pm \sqrt{N-1}$. $\alpha$ and $\gamma$ take the values
$\alpha$=$\pm \sqrt{\frac{1}{2}\frac{(N-1)}{(N-2)}}$ and $\gamma$=$\alpha
+\frac{N-4}{N-3}(1-\alpha )$.\ \ $\eta $ is an arbitrary, positive
constant of dimension ${\mathrm{length}}^{-2}$. The massless scalar
field is\ \
\begin{equation}
\phi =\sigma  V( x) +\rho  \sqrt{\eta }Y ,
\end{equation}where $\sigma \equiv \pm \sqrt{\frac{(N-2)}{\kappa
( N-3)  } (1-\alpha ^{2})}=\pm \frac{1}{\sqrt{2 \kappa }} $and $\rho
\equiv \sqrt{\frac{1}{\kappa }(N-2)}\lambda $.
\end{corollary}

\subsubsection{Class $\mathcal{T}_{N}^{0}$ solutions with non-vanishing
scalar potential}

Next, let us consider the class of $\mathcal{T}_{N}^{0}$solutions
with non-vanishing scalar field potential; $W_{0}\neq 0$. These
solutions can be inferred directly from Theorem \ref{XRef-Theorem-103203143}
and eq. \ref{XRef-Equation-5465721}. In 4 dimensions and for static
vacuum solutions, they are the solutions found by Fonarev in \cite{Fonarev:1994xq}.\ \ According
to eq. \ref{XRef-Equation-5465721}, the metric function $\mu$ is
in this case given by
\[
\mu =\begin{cases}
\sqrt{\eta  }y & \eta  >0 ,\lambda ^{2}=1 \\
\frac{1}{\lambda ^{2}-1}\ln |\sqrt{\eta  }\left( \lambda ^{2}-1\right)
y| &  \eta  > 0, \lambda ^{2}\neq 1 \\
\end{cases}
\]

For the special case $\lambda ^{2}=1$, the solution takes the form
\begin{gather}
{\mathrm{ds}}^{2}= \epsilon  e^{2\left( \alpha  V+\sqrt{\eta  }y\right)
}{\mathrm{dy}}^{2}+e^{-2\left( \gamma  V-\sqrt{\eta  }y\right) }{\hat{h}}_{\mathrm{ij}}{\mathrm{dx}}^{i}{\mathrm{dx}}^{j}
\\\phi =\sigma  V+\sqrt{\frac{\left( N-2\right) }{\kappa }}\sqrt{\eta
}y,
\end{gather}

In the general case for $\lambda ^{2}\neq 1$, we do a coordinate
transformation to a new coordinate $Y$ defined by $\ln |(\lambda
^{2}-1)\sqrt{\eta  } y|=(\lambda ^{2}-1)\sqrt{\eta  } Y$. We now
get the metric on the form
\[
{\mathrm{ds}}^{2}= \epsilon  e^{2\left( \alpha  V+\lambda ^{2}\sqrt{\eta
}Y\right) }{\mathrm{dY}}^{2}+e^{-2\gamma  V+2\sqrt{\eta  }Y}{\hat{h}}_{\mathrm{ij}}{\mathrm{dx}}^{i}{\mathrm{dx}}^{j}
\]

and the scalar field takes the form
\[
\phi =\sqrt{\frac{\left( N-2\right) }{\kappa }}\left( \varepsilon
\frac{1}{\sqrt{\left( N-3+\lambda ^{2}\right) }}V+\lambda \sqrt{\eta
} Y\right)
\]

We see that the metric and scalar field for $\lambda ^{2}\neq 1$
coincide with those for the case $\lambda ^{2}=1$. Thus, we can
summarize this result as follows:
\begin{corollary}

 $(\mathcal{T}_{N}^{0}, W_{0}\neq 0)$ The $\mathcal{T}_{N}^{0}$class
of solutions with non-vanishing scalar field potential is a two-parameter
family of solutions given by the metric \label{XRef-Corollary-109203724}
\begin{equation}
{\mathrm{ds}}^{2}= \epsilon  e^{2\left( \alpha  V+\lambda ^{2}\sqrt{\eta
}Y\right) }{\mathrm{dY}}^{2}+e^{-2\gamma  V+2\sqrt{\eta  }Y}{\hat{h}}_{\mathrm{ij}}{\mathrm{dx}}^{i}{\mathrm{dx}}^{j},
\end{equation}a scalar field on the form
\begin{equation}
\phi =\sqrt{\frac{\left( N-2\right) }{\kappa }}\left( \varepsilon
\frac{1}{\sqrt{\left( N-3+\lambda ^{2}\right) }}V+\lambda \sqrt{\eta
} Y\right)
\end{equation}and a scalar field potential
\begin{equation}
W[ \phi ] =W_{0}e^{-2k \phi },
\end{equation}where k=$\sqrt{\frac{\kappa }{N-2}}\lambda $ and $W_{0}=-\epsilon
\frac{(N-2)}{2\kappa  }(N-1-\lambda ^{2})\eta $, $\eta$ > 0.
\end{corollary}

Corollary \ref{XRef-Corollary-101818159} in combination with Corollary
\ref{XRef-Corollary-109203724} generalizes the Fonarev result \cite{Fonarev:1994xq}
to arbitrary dimension and for arbitrary orthogonal-symmetric seed
metrics. We also see that the solutions of corollaries \ref{XRef-Corollary-101818159}
and \ref{XRef-Corollary-109203724} have the same form, so Corollary
\ref{XRef-Corollary-109203724} can be trivially extended to also
cover the solutions of Corollary \ref{XRef-Corollary-101818159}
by\ \ allowing parameter values $\eta =0$ and\ \ $\lambda =\pm \sqrt{N-1}$.

\subsubsection{Class $\mathcal{T}_{N}^{\Lambda }$ solutions with
vanishing scalar potential}

Now, turning our attention to the classes of solutions with $\Lambda
\neq 0$, the $\mathcal{T}_{N}^{\Lambda }$class of scalar field solutions,
we begin by investigating the case of solutions with vanishing scalar
field potential, which requires that $W_{0}=0$. The constraints
that must be satisfied in this case are listed in Theorem \ref{XRef-Theorem-103203143},
and are equations \ref{XRef-Equation-10322143}, \ref{XRef-Equation-10321522},
\ref{XRef-Equation-103214959} and \ref{XRef-Equation-10982211}.
The solution to these constraints is given by a discrete set of
parametric values corresponding to the value $\lambda =$$\varepsilon
/\sqrt{N-1}$:
\[
\alpha =\frac{1}{N-2}, \sigma =\varepsilon \sqrt{\frac{N-1}{\kappa
( N-2) }},\ \ \rho =\varepsilon \sqrt{\frac{N-2}{\kappa ( N-1) }},
\]

where $\varepsilon  = \pm 1$. According to eq. \ref{XRef-Equation-5465721},
the metric function $\mu $ takes the form $\mu =\sqrt{-\chi }y$,
which requires $\chi  < 0$. Let us summarize this solution:
\begin{corollary}

 $(\mathcal{T}_{N}^{\Lambda }, W_{0}=0)$ The set of $\mathcal{T}_{N}^{\Lambda
}$ solutions with vanishing scalar field potential is a one-parameter
family of solutions parametrized by the parameter $\chi $ with $\chi
< 0$. The solutions correspond to a choice of cosmological constant
in the seed solution satisfying\ \ $\Lambda =-\frac{\epsilon }{2}\frac{{(N-2)}^{3}}{(N-1)}\chi
$.\ \ The other parameters are defined by equations \ref{XRef-Equation-54222627}
- \ref{XRef-Equation-55165629}\ref{XRef-Equation-54222627}with $\lambda
= \varepsilon /\sqrt{N-1}$, and consequently $\alpha =1/(N-2)$.
The metric is given by \label{XRef-Corollary-101818412}
\begin{equation}
{\mathrm{ds}}^{2}= \epsilon  e^{2\left( \alpha  V+\sqrt{-\chi }y\right)
}{\mathrm{dy}}^{2}+e^{-2\gamma  V+2\sqrt{-\chi }y}{\hat{h}}_{\mathrm{ij}}{\mathrm{dx}}^{i}{\mathrm{dx}}^{j},
\end{equation}with a scalar field
\begin{equation}
\phi =\varepsilon \sqrt{\frac{N-1}{\kappa ( N-2) }}\left( V+\frac{\left(
N-2\right) }{\left( N-1\right) }\sqrt{-\chi }y\right) .
\end{equation}$W[ \phi ]  = 0$
\end{corollary}

\subsubsection{Class $\mathcal{T}_{N}^{\Lambda }$ solutions with
non-vanishing scalar potential}

The final part of the $\mathcal{T}_{N}$solution space to be analyzed
is the $\mathcal{T}_{N}^{\Lambda }$ class of solutions with non-vanishing
scalar field potential. Proceeding as before, we first identify
the constraints to solve. In this case it is the full set of constraints,
namely equations \ref{XRef-Equation-10322143}, \ref{XRef-Equation-10321522},
\ref{XRef-Equation-1098246}, \ref{XRef-Equation-103214959}, \ref{XRef-Equation-10982211}
and \ref{XRef-Equation-103193044}.\ \ There are two unique solutions
to these constraints:
\[
\alpha =\frac{1}{N-2}, \sigma =\varepsilon \sqrt{\frac{N-1}{\kappa
( N-2) }},\ \ \rho =\varepsilon \sqrt{\frac{N-2}{\kappa ( N-1) }},
k=\varepsilon \sqrt{\frac{\kappa }{\left( N-1\right) N-2}},
\]

where $\varepsilon  = \pm 1$, as before. These parameter values
correspond to $\lambda  = \varepsilon  /\sqrt{N-1}$. The expression
for $\mu $ is given by eq. \ref{XRef-Equation-5465721}. By transforming
to new coordinates, we are able to simplify the expressions considerably.
Define a new coordinate $\tau $ by
\[
\tau ^{\left( N-2\right) }=\begin{cases}
|\sqrt{\frac{\chi }{\eta }}{\cos ( \left( 1-\lambda ^{2}\right)
\sqrt{\chi }y) }^{-1}| & \left( \frac{\eta }{\chi }>0, \chi \neq
0, \lambda ^{2}\neq 1\right)  \\
|\sqrt{\frac{\chi }{\eta }}{\sin ( \left( 1-\lambda ^{2}\right)
\sqrt{\chi }y) }^{-1}| & \left( \eta >0, \chi <0, \lambda ^{2}\neq
1\right)  \\
\end{cases}
\]

Doing so, we are able to simplify the expressions considerably.
Let us summarize the result:
\begin{corollary}

 $(\mathcal{T}_{N}^{\Lambda }, W_{0}\neq 0)$ The $\mathcal{T}_{N}^{\Lambda
}$class of solutions with non-vanishing scalar field potential is
a two-parameter family of solutions given by the metric \label{XRef-Corollary-1018183318}
\begin{equation}
{\mathrm{ds}}^{2}=\epsilon  e^{2\alpha  V}\frac{{\left( N-1\right)
}^{2}}{\left( \eta -\chi  \tau ^{-2\left( N-2\right) }\right) }{\mathrm{d\tau
}}^{2}+{\tau }^{2\left( N-1\right) }e^{-2\gamma  V}{\hat{h}}_{\mathrm{ij}}{\mathrm{dx}}^{i}{\mathrm{dx}}^{j},
\end{equation}a scalar field on the form
\begin{equation}
\phi =\varepsilon \sqrt{\frac{N-1}{\kappa ( N-2) }} \left( V + \left(
N-2\right) \ln  \tau \right)
\end{equation}and a scalar field potential
\begin{equation}
W[ \phi ] =W_{0}e^{-2k \phi },
\end{equation}where $\alpha =1/(N-2), k=\sqrt{\frac{\kappa }{(N-1)(N-2)}}$
and\ \ $W_{0}=-\epsilon  \frac{(N-2)}{2\kappa }(N-1-\lambda ^{2})
\eta $, corresponding to $\lambda  = \varepsilon  /\sqrt{N-1}$.
$\eta $ and $\chi $ are free parameters of dimension ${\mathrm{length}}^{-2}$.
The $\mathcal{T}_{N}^{\Lambda }$solutions must have $\chi  \neq
0$ and satisfy at least one of the conditions $\eta$ > 0 or $\chi$
< 0.
\end{corollary}

\subsection{General form of the solutions}

We will now seek a common form of all class $ \mathcal{S}$ and $
\mathcal{T}$ solutions that is capable of representing all solutions.
In order to do so, we must introduce yet another set of parameters.
We will continue to use the parameter $\lambda$ used before, but
define the new parameters $q\equiv \sqrt{\eta -\chi }$ and $b\equiv
\sqrt{\frac{\eta }{\eta -\chi }}$. Writing $\eta$ and $\chi$ in
terms of $q$ and $b$ gives\ \ $\eta =q^{2}b^{2}$ and $\chi =(b^{2}-1)q^{2}$.\ \

It is now straight forward to prove our main result in terms of
parameters $\lambda $, $q$ and $b$, which we summarize in the form
of a new theorem:
\begin{theorem}

Given an orthogonal-symmetric vacuum solution, $\hat{g}=$g[V,$\hat{h}$],
to the N dimensional Einstein equations with an arbitrary cosmological
constant $\Lambda$ (equation \ref{XRef-Equation-92783053}),\ \ the
complete set of class $ \mathcal{S}$ and $ \mathcal{T}$ solutions
derived from $\hat{g}$ has a space-time metric \label{XRef-Theorem-101785048}
\begin{equation}
{\mathrm{ds}}^{2}=\epsilon \frac{e^{2\left( \alpha  V+q Y\right)
}}{b^{2} e^{2\left( 1-\lambda ^{2}\right) q Y }+1-b^{2}}{\mathrm{dY}}^{2}+e^{-2\gamma
V+2q Y}{\hat{h}}_{\mathrm{ij}}{\mathrm{dx}}^{i}{\mathrm{dx}}^{j}%
\label{XRef-Equation-101817047}
\end{equation}and a scalar field
\begin{equation}
\phi =\sqrt{\frac{\left( N-2\right) }{\kappa }}\left( \varepsilon
\frac{1}{\sqrt{\left( N-3+\lambda ^{2}\right) }}V+\lambda  q Y\right)
\label{XRef-Equation-10181714}
\end{equation}with a scalar field potential
\begin{equation}
W[ \phi ] =W_{0}e^{-2k \phi },%
\label{XRef-Equation-101785254}
\end{equation}where $\varepsilon  = \pm 1$, $\lambda$ and b are
dimensionless parameters and q is a parameter of dimension inverse
length. Both b and q must be $ \geq $ 0. The constants $\alpha$,
$\gamma$, k and $W_{0}$ are given by
\begin{equation}
\alpha =\varepsilon \frac{\lambda }{\sqrt{N-3+\lambda ^{2}}}, k=\sqrt{\frac{\kappa
}{N-2}}\lambda , W_{0}=-\epsilon  \frac{\left( N-2\right) \left(
N-1-\lambda ^{2}\right) }{2\ \ \kappa } q^{2}b^{2}, %
\label{XRef-Equation-1018173355}
\end{equation}and the parameters q and b are related to $\Lambda$
as follows:
\begin{equation}
\Lambda =-\epsilon  \frac{\left( N-2\right) \left( N-3 \right) \left(
N-1-\lambda ^{2}\right) }{2\ \ \left( N-1-2 \alpha \right) }q^{2}
\left( b^{2}-1\right) .%
\label{XRef-Equation-101817345}
\end{equation}The main solution classes can be retrieved from equations
\ref{XRef-Equation-101817047} and \ref{XRef-Equation-10181714} as
follows:
\begin{itemize}
\item $\mathcal{S}_{N}$: The complete set of class $\mathcal{S}_{N}$solutions
is obtained from equations \ref{XRef-Equation-101817047} and \ref{XRef-Equation-10181714}
when $q=0$ and $\lambda$ $\epsilon$ $(-\infty , \infty )$.
\item $\mathcal{T}_{N}^{0}$: The $\mathcal{T}_{N}^{0}$ solutions
are obtained from equations \ref{XRef-Equation-101817047} and \ref{XRef-Equation-10181714}
when $b=1$, $q\ \ > 0$ and $\lambda$ $\epsilon$ $(-\infty , \infty
)$.
\item $\mathcal{T}_{N}^{\Lambda }$: The $\mathcal{T}_{N}^{\Lambda
}$solutions are obtained from equations \ref{XRef-Equation-101817047}
and \ref{XRef-Equation-10181714} when $b\neq 1$, $q\ \ > 0$ and
$\lambda  =\varepsilon /\sqrt{N-1}$.
\end{itemize}
\end{theorem}
\begin{proof}

If we first consider the case of the empty $\mathcal{S}_{N}^{\Lambda
}$ solution class of Corollary \ref{XRef-Corollary-1018173233},
we see that the fact that there are no solutions of this class is
reflected by equations \ref{XRef-Equation-1018173355} and \ref{XRef-Equation-101817345},
because both $W_{0}\neq 0$ and $\Lambda \neq 0$ require $q\neq 0$.
From the metric in equation \ref{XRef-Equation-101817047} we see
that this means that a class $ \mathcal{T}$ solution is generated,
thus there are no solutions of class $\mathcal{S}_{N}^{\Lambda }$.
Next, Corollary \ref{XRef-Corollary-104611} follows directly from
equations \ref{XRef-Equation-101817047} and \ref{XRef-Equation-10181714}
when setting $q = 0$. Corollary \ref{XRef-Corollary-101818159} follows
directly from equations \ref{XRef-Equation-101817047} and \ref{XRef-Equation-10181714}
when we let $q > 0$, $\lambda  = \pm  \sqrt{N-1}$ and\ref{XRef-Corollary-101818159}$b=1$\ref{XRef-Corollary-101818159}
Similarly, Corollary \ref{XRef-Corollary-109203724} can be inferred
from equations \ref{XRef-Equation-101817047} and \ref{XRef-Equation-10181714}
by letting $q > 0$ and setting $b=1$. Corollary \ref{XRef-Corollary-101818412}
follows from equations \ref{XRef-Equation-101817047} and \ref{XRef-Equation-10181714}
by letting $q^{2}=-\chi $,\ \ $\lambda =\pm 1/\sqrt{N-1}$, $q >
0$ and $b=0$. Finally, the results of Corollary \ref{XRef-Corollary-1018183318}
can be brought to the form of equations \ref{XRef-Equation-101817047}
and \ref{XRef-Equation-10181714} by transforming to a new coordinate
$Y$ defined by $\tau ^{(N-1)}= e^{\sqrt{\eta -\chi }Y}.$ The completeness
of theorem \ref{XRef-Theorem-101785048} follows from the way we
derived it: a general form of the potential was investigated, from
which we derived the potential of equation \ref{XRef-Equation-101785254}.
Furthermore, we did an exhaustive analyzis of all options for satisfying
the constraints set up by Theorem \ref{XRef-Theorem-103203143},
which represent all class $ \mathcal{T}$ and $ \mathcal{S}$ solutions.
\end{proof}

\section{Example solutions}\label{XRef-Section-5159268}

No solutions within the $\mathcal{T}_{N}^{\Lambda }$ class of solutions
have been published before, so let us have a closer look at a particular
solution of this class in 4 dimensions. As the class requires seed
solutions with non-zero cosmological constant, let us use the (Anti-)
deSitter solution in 4 dimensions as our seed solution. Take $\Lambda
$ as a non-zero cosmological constant. The $(A){\mathrm{dS}}_{4}$
metric can be cast into the following static form:
\begin{equation}
{\mathrm{ds}}^{2}=-\left( 1-\frac{\Lambda  }{3}r^{2}\right) {\mathrm{dt}}^{2}+{\left(
1-\frac{\Lambda  }{3}r^{2}\right) }^{-1}\left( {\mathrm{dr}}^{2}+\left(
1-\frac{\Lambda  }{3}r^{2}\right) r^{2}{\mathrm{d\Omega }}^{2}\right)
.%
\label{XRef-Equation-526103121}
\end{equation}

\subsection{Minimally coupled solution}

Define $\ell ^{2}\equiv 3/|\Lambda |$. Applying Corollary \ref{XRef-Corollary-1018183318}
and defining a new coordinate
\[
\psi \equiv \begin{cases}
\arcsin ( r/\ell )  & \left( \Lambda  > 0\right)  \\
\operatorname{arcsinh}( r/\ell )  & \left( \Lambda  < 0\right)
\\
\end{cases}
\]

 allows us to cast the solution metric into the following form:
\begin{equation}
{\mathrm{ds}}^{2}= 9\eta ^{-1}C( \psi ) \left( -\frac{ {\mathrm{d\tau
}}^{2}}{\left( 1-\frac{\chi }{\eta } \tau ^{-4}\right) }+\tau ^{6}\frac{\eta
}{4\chi } \left( {\mathrm{d\psi }}^{2}+S^{2}( \psi )  {\mathrm{d\Omega
}}^{2}\right) \right) %
\label{XRef-Equation-5106438}
\end{equation}

where $S( \psi ) \equiv  \sin  \psi $ for $\Lambda >0$, while $S(
\psi ) \equiv \sinh  \psi $ when $\Lambda  <0$. Likewise, $C( \psi
) \equiv \cos  \psi $ if $\Lambda >0$, while $C( \psi ) =\cosh
\psi $ when $\Lambda <0$. In the case of positive $\Lambda $, the
spatial slices of this geometry are conformally equivalent to the
hypersphere $S^{3}$ and the solution topology is $R\times S^{3}$,
while for negative $\Lambda $, the spatial slices are conformally
equivalent to\ \ the hyperboloid $H^{3}$ and the solution topology
is $R\times H^{3}$.\ \ The scalar curvature is
\begin{equation}
R=\begin{cases}
\frac{2 \eta  \left( 15 \tau ^{4} {\cos }^{2}\psi +\frac{\chi }{\eta
} \right) }{3 \tau ^{6} {\cos }^{3} \psi } & \left( \Lambda >0\right)
\\
-\frac{2\eta \ \ \ \left( \left( 18 \frac{\chi }{\eta }-15 \tau
^{4}\right) {\cos }^{2}\psi +\frac{\chi }{\eta } \right) }{3 \tau
^{6}{\cosh }^{3}\psi } & \left( \Lambda <0\right)  \\
\end{cases}%
\label{XRef-Equation-51064940}
\end{equation}

The scalar field takes the unique form
\begin{equation}
\phi =-\sqrt{\frac{3}{2\kappa }} \ln (  \tau ^{2} C( \psi )  ) ,%
\label{XRef-Equation-511004}
\end{equation}

which is independent of parameters $\eta $ and $\chi$. The metric
of eq. \ref{XRef-Equation-5106438} is singular for time $\tau ^{4}=\chi
/\eta $, while the Ricci scalar of eq. \ref{XRef-Equation-51064940}
is regular. By transforming the Riemann tensor to an orthonormal
basis, we find that $\tau ^{4}=\chi /\eta $ is just a coordinate
singularity.

The geometry has an initial singularity at $\tau =0$, and there
is a curvature singularity at $\psi =\pm \pi /2$ for $\Lambda  >
0$. As $\psi \rightarrow \pm \pi /2$, the scalar field $\phi \rightarrow
+\infty $ . The parameters $\eta $ and $\chi $ evaluate to $\eta
=\frac{3}{8}\kappa  W_{0}$ and $\chi =\frac{3}{4}\Lambda =\frac{9}{4\ell
^{2}}$, so $\eta /\chi =\kappa  W_{0}/2\Lambda $.

\subsection{Conformally coupled solution}

We will now transform the de Sitter-based solution we generated
in the previous section to the conformal frame, using the generalized
Bekenstein transformation for arbitrary scalar potential \cite{Maeda:1988ab}.
The generalized Bekenstein transformation is described in Theorem
\ref{XRef-Theorem-510234826} of Appendix \ref{XRef-AppendixSection-510235039}.
Applying it to the solution of eqs. \ref{XRef-Equation-5106438}
and \ref{XRef-Equation-511004} for $\Lambda >0$ gives a metric of
the transformed solution:
\begin{equation}
{\mathrm{ds}}^{2}={\left( \cos  \psi +\frac{1}{2T}\right) }^{2}\frac{
9}{4}\eta ^{-1}\left( -\frac{ dT^{2}}{\left( 1-\frac{\chi }{4\eta
} T^{-2}\right) }+T^{4}\frac{4\eta }{\chi } \left( {\mathrm{d\psi
}}^{2}+{\sin }^{2}\psi  {\mathrm{d\Omega }}^{2}\right) \right) ,
\end{equation}

where $0\leq \psi \leq \pi $, and we have introduced a new time
coordinate $T\equiv \frac{1}{2}\tau ^{2}$. The conformally coupled
scalar field $\Phi $ is
\begin{equation}
\Phi =\frac{1}{\zeta }\tanh  \zeta  \phi =-\frac{1}{\zeta }\frac{\left(
2T \cos  \psi -1\right) }{\left( 2T \cos  \psi +1\right) },
\end{equation}

with $\zeta =\sqrt{\kappa /6}$. The scalar potential of this solution
is
\begin{equation}
W[ \Phi ] ={W_{0}( 1-\zeta ^{2}\Phi ^{2}) }^{2}e^{-2\zeta  \phi
}=W_{0}( 1-\zeta ^{2}\Phi ^{2}) {\left( 1-\zeta  \Phi \right) }^{2}.
\end{equation}

The Ricci scalar of this geometry is
\begin{equation}
R=\frac{192 \left( 1+6 T \cos  \psi \right) }{ 9 {\left( 1+2 T \cos
\psi \right) }^{3}}\eta .%
\label{XRef-Equation-5129038}
\end{equation}

Thus, there is a curvature singularity where $\cos  \psi =-1/2T$.
Thus, for $|T|<\frac{1}{2}$, the geometry is regular everywhere.
By transforming the Riemann tensor to an orthonormal basis, we find
that the solution is free of other singularities. Introduce constants
$H_{0}\equiv \frac{4}{3}\sqrt{\eta }$ and $\ell \equiv \frac{3}{2\sqrt{\chi
}}=\sqrt{3/\Lambda }$. Then define a new time coordinate $t$ by
$t\equiv \epsilon  \ell \sqrt{T^{2}-\frac{\chi }{4\eta  }}$, where
$\epsilon \equiv \mathrm{sign} T$. We are now able to cast the metric
into the following form:
\begin{equation}
{\mathrm{ds}}^{2}={\left( \frac{1}{\sqrt{a( t) }}+c \cos  \psi \right)
}^{2}\left( - {\mathrm{dt}}^{2}+{a( t) }^{2}H_{0}^{-2} \left( {\mathrm{d\psi
}}^{2}+{\sin }^{2}\psi  {\mathrm{d\Omega }}^{2}\right) \right) ,%
\label{XRef-Equation-513141035}
\end{equation}

where $c$ is a dimensionless constant, $c\equiv 2{(\ell  H_{0})}^{-1}=\sqrt{\chi
/\eta }$, and $a( t) $ is a time-dependent scale factor; $a( t)
\equiv H_{0}^{2}t^{2}+1$. The cosmological time scale is $H_{0}^{-1}$.
Similarly, $\ell  $ is a cosmological length scale that is identical
to the length scale of the ${\mathrm{dS}}_{4}$ seed metric, eq.
\ref{XRef-Equation-526103121}. We can then write the ratio $\eta
/\chi $ in terms of $\ell $ and $H_{0}$: $\frac{\eta }{\chi }=\frac{1}{4}\ell
^{2}H_{0}^{2}$. Notice that $c^{2}$ is proporional to $\Lambda /\kappa
W_{0}$, so $c$ is defined by the relative size of the cosmological
constant over the strength of the scalar potential. In the limit
$c\rightarrow 0$, the geometry of eq. \ref{XRef-Equation-513141035}
is spatially homogenous. We can therefore refer to $c$ as the {\itshape
inhomogeneity parameter} of the spacetime. A cosmological constant
that is weak relative to the strength of the scalar potential yields
a geometry that is highly homogeneous (low $c$), while a strong
cosmological constant yields a highly inhomogenous cosmology (high
$c$). The scale factor $a( t) $ is $1$ for $t=0$. This\ \ solution
is clearly a spatially finite, inhomogeneous, inflationary cosmology
with a time scale set by $H_{0}$ and a size of the spatial sections
governed by the length scale $\ell ={(c H_{0})}^{-1}$ and the scale
factor $a( t)  $. Corresponding to times $T=\pm 1/2$, this universe
has an initial curvature singularity that vanishes at time $t_{-}=-H_{0}^{-1}\sqrt{\frac{1}{c}-1}$
and a final curvature singularity that starts to emerge from $\psi
=\pi $ at time $t_{+}=H_{0}^{-1}\sqrt{\frac{1}{c}-1}$. From eq.
\ref{XRef-Equation-5129038} we see that the geometry has positive
curvature everywhere. It has a finite volume of
\begin{equation}
{\mathrm{vol}}_{\mathcal{M}}=\frac{\pi ^{2}}{2} \left( 4+3 a( t)
c^{2}\right) \ \ a{\left( t\right) }^{3/2}H_{0}^{-3}
\end{equation}

Thus, the size of the cosmology is defined by the time constant
$H_{0}^{-1}$, the scale factor $a( t) $, as well as the inhomogeneity
$c$. A weak scalar potential implies a long time scale and a small
cosmological constant implies a cosmology of large size. Furthermore,
if $c<<1$, the time span between the initial and final singularities
are much larger than the cosmological time scale: $c<<1\Rightarrow
|t_{+}-t_{-}|>>H_{0}^{-1}$.

Looking at asymptotic behaviors of this geometry, we find that for
$|t|<<H_{0}^{-1}$, the geometry is nearly static because, at $t=0$,
$\dot{a}=0$ and $a( t) \simeq 1$:
\[
{\mathrm{ds}}^{2}\simeq {\left( 1+c \cos  \psi \right) }^{2}\left(
- {\mathrm{dt}}^{2}+ H_{0}^{-2}( {\mathrm{d\psi }}^{2}+{\sin }^{2}(
\psi )  {\mathrm{d\Omega }}^{2}) \right)
\]

 This means that the size of the cosmology is nearly constant at
times $|t|<<H_{0}^{-1}$; in fact $\dot{a}=0$ for $t=0$, but the
geometry is still accelerating.

At late times, $t >>H_{0}^{-1}$, the cosmology approaches power-law
inflation:
\[
{\mathrm{ds}}^{2}\simeq c^{2} {\cos }^{2}\psi  \left( -{\mathrm{dt}}^{2}+{\left(
H_{0}t\right) }^{4}H_{0}^{-2} \left( {\mathrm{d\psi }}^{2}+{\sin
}^{2}\psi  {\mathrm{d\Omega }}^{2}\right) \right) .
\]

Now, if we define a new time coordinate $z\equiv \arctan ( H_{0}t)
$, the metric of eq. \ref{XRef-Equation-513141035} takes the form
\begin{equation}
{\mathrm{ds}}^{2}={\left( \frac{\cos  z+c \cos  \psi }{H_{0}{\cos
}^{2}z}\right) }^{2} \left( -{\mathrm{dz}}^{2}+{\mathrm{d\psi }}^{2}+{\sin
}^{2}\psi  {\mathrm{d\Omega }}^{2}\right) ,%
\label{XRef-Equation-51318159}
\end{equation}

where the conformal time $z$ is defined in the interval $-\pi /2<z<\pi
/2$. Thus, the solution geometry $\mathcal{M}$ is conformally isometric
to the Einstein static universe $\tilde{\mathcal{M}}=R\times S^{3}$,
and maps onto the $-\pi /2<z<\pi /2$ section of the Einstein cylinder.
Then, $t\rightarrow \pm \infty $ as $z\rightarrow \pm \pi /2$. The
conformal boundary of the spacetime consists of the two timelike
conformal infinities $i^{\pm }$ at $z=\pm \pi /2$, respectively.
All timelike geodesics emerge from $i^{-}$ and end at $i^{+}$. If
we analyze null geodesics on $\tilde{\mathcal{M}}$, we find that
null geodesics in the $\psi$ direction satisfy $\psi -\psi _{0}=\pm
z$, where $\psi _{0}\equiv \psi ( z=0) $. Thus, due to rotational
symmetry of $S^{3}$, photons can only travel an angular distance
$< \pi /2$ in a finite amount of time in any spatial direction of
$\tilde{\mathcal{M}}$. Subsequently, for a stationary observer at
an arbitrary spatial location, there is a cosmological event horizon
at an angular distance $\pi /2$ away from the observer, implying
that the observable universe is half the total universe. If we transform
the cosmological horizon back to the physical solution geometry
of $\mathcal{M}$, the horizon will be at the same spatial coordinates
on $\mathcal{M}$ as on $\tilde{\mathcal{M}}$, but the geometry of
the horizon will be changed by the conformal factor of eq. \ref{XRef-Equation-51318159}.
Still, the observable universe extends an angular distance of $\pi
/2$ away from the stationary observer.

\begin{figure}[h]
\begin{center}
\includegraphics{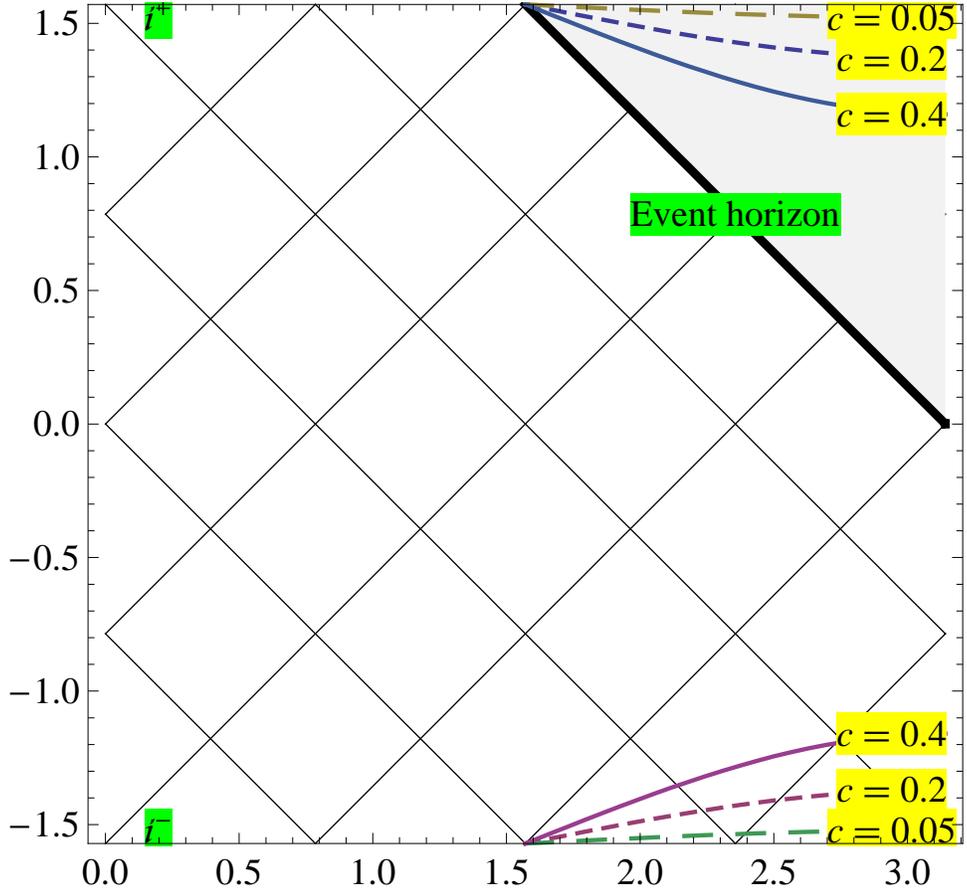}

\end{center}
\caption{Penrose-Carter diagram for $\psi$-directed null geodesics.
Initial and final singularities are plotted for $c=0.05$ (long dash),
$c=0.2$ (short dash), $c=0.4$ (continuous). The black hole event
horizon is the $z=\pi -\psi $ curve, and the gray region $z+\psi
> \pi $ is the black hole region for all solutions with $c < 1$.}
\label{XRef-FigureCaption-5307522}
\end{figure}

Figure \ref{XRef-FigureCaption-5307522} shows a Penrose-Carter diagram
for this spacetime. We see from this diagram that $\psi$-directed
null geodesics, with the exception of the points on the $\psi =\pi
/2$ surface, either emerge from the initial singularity or terminate
at the final singularity. Furthermore, there are timelike geodesics
that do not intersect any of the singularities. We can also see
that in the region $z+\psi >\pi $, all $\psi$-directed geodesics
that are either null or time-like intersect the final singularity.
Furthermore, we can infer from the\ \ Penrose-Carter diagram that
for all solutions with $c<1$, the null surface $z+\psi =\pi $ constitutes
an event horizon, because no null or time-like geodesics with $z+\psi
>\pi $ will ever enter the $z+\psi <\pi $ region. The solution is
regular at the $z+\psi =\pi \text{}$ surface. From the metric of
eq. \ref{XRef-Equation-51318159}, we see that the initial and final
singularities are in fact point singularities, because the 2 dimensional
surfaces $(z=\mathrm{const}, \psi =\mathrm{const})$ have vanishing
proper surface area as $\cos  z+c \cos  \psi \rightarrow 0$.\ \ We
can conclude that the $c<1$ solutions are solutions in which a scalar
black hole emerges. The event horizon emerges at conformal time
$z=0$, whereas the curvature singularity emerges later, at conformal
time $z_{+}=\arctan ( H_{0}t_{+}) =\arctan ( \sqrt{\frac{1}{c}-1})
$.\ \

In the terms of the conformal time $z$, the scalar field takes the
following form:
\begin{equation}
\Phi =-\frac{1}{\zeta }\frac{\left( c\ \ \cos  \psi -\cos  z\right)
}{\left( c\ \ \cos  \psi +\cos  z\right) }%
\label{XRef-Equation-514234249}
\end{equation}

We see from eq. \ref{XRef-Equation-514234249} that the scalar field
blows up at the singularity, but is regular everywhere else, including
at the event horizon.

Let us look at energy conditions. The weak and strong energy conditions
can be expressed in terms of the energy density $\rho $ and the
principal pressures $p_{i}$ evaluated in the rest frame of the spacetime
fluid \cite{Wald:1984-general}. The weak energy condition requires
$\rho \geq 0$ and $\rho +p_{i}\geq 0$, while the strong energy condition
requires $\rho +p_{i}\geq 0$ and $\rho +\Sigma _{i}p_{i}\geq 0$.
Using the Einstein equations, we may evaluate $\rho $ and $p_{i}$
from the Einstein tensor:
\[
 \begin{array}{cc}
 \rho =-\kappa ^{-1}G_{0}^{0} &   \\
 p_{i}=\kappa ^{-1}G_{i}^{i} & \left( i=1,2,3\right)
\end{array}
\]

For the metric of eq. \ref{XRef-Equation-51318159}, this gives
\begin{gather}
\rho =H_{0}^{2}\frac{ {\cos }^{2}z\ \ \left( 3+c^{2}( 11 -\cos
2z) +3\cos  2z+24 c\ \ \cos  z \cos  \psi +2 c^{2} \left( 5-\cos
2z \right)  \cos  2\psi \right) }{2 {\left( \cos  z+c \cos  \psi
\right) }^{4}}%
\label{XRef-Equation-514215635}
\\\rho +p_{\psi }=H_{0}^{2}\frac{2 c^{2}{\cos }^{2}z\ \ \ \left(
1-\cos  2z \cos  2\psi \right) }{ {\left( \cos  z+c \cos  \psi \right)
}^{4}}%
\label{XRef-Equation-514215836}
\\\rho +p_{\theta }=\rho +p_{\varphi }=H_{0}^{2}\frac{c^{2} {\cos
}^{2}\psi  {\sin }^{2}2z}{ {\left( \cos  z+c \cos  \psi \right)
}^{4}}%
\label{XRef-Equation-514215856}
\end{gather}

From eq. \ref{XRef-Equation-514215635} it follows immediately\ \ that
the energy density is positive definite only for small $c$ ($c<1/8$)
and large c ($c>24$). Furthermore, eqs. \ref{XRef-Equation-514215836}
and \ref{XRef-Equation-514215856} yield $\rho +p_{i}\geq 0$ on all
of $\mathcal{M}$. Thus, the scalar field satisifes the weak energy
condition on all of $\mathcal{M}$ for $c<1/8$ or $c>24$. Furthermore,
\[
\rho +\Sigma _{i}p_{i}=-H_{0}^{2}\frac{ {\cos }^{2}z\ \ \left( 3\left(
1+\cos  2z\right) +24 c \cos  z \cos  \psi + c^{2}( 7+ \cos  2z+8\cos
2\psi +2\cos  2z \cos  2\psi )  \right) }{{\left( \cos  z+c \cos
\psi \right) }^{4}}
\]

 $\rho +\Sigma _{i}p_{i}$ may take both positive and negative values
on $ \mathcal{M}$, so the strong energy condition is generally violated
on $\mathcal{M}$.

Finally, let us have a look at the equation of state parameter $w_{i}\equiv
p_{i}/\rho $. Defining $u\equiv {H_{0}^{2}( \cos  z+c \cos  \psi
) }^{-4}$, eqs. \ref{XRef-Equation-514215836} and \ref{XRef-Equation-514215856}\ \ allow
us to write
\begin{gather}
w_{\psi }=-1+2 c^{2}{\cos }^{2}z\ \ \ \left( 1-\cos  2z \cos  2\psi
\right) \frac{u}{\rho }
\\w_{\theta }=w_{\varphi }=-1+c^{2}( {\cos }^{2}\psi  {\sin }^{2}2z
) \frac{u}{\rho }
\end{gather}

 $w_{i}>-1$ when $\rho >0$. Thus, the weak energy condition is satisfied
when $\rho >0$. We also see that $w_{i}\rightarrow -1$ as $c\rightarrow
0$. This happens when the cosmological constant is much weaker than
the strength of the scalar potential. In that case, we saw from
eq. \ref{XRef-Equation-514234249} that the scalar field is nearly
constant and approaches the Planckian value $\zeta ^{-1}$.
\begin{figure}[h]
\begin{center}
\includegraphics[width=150mm]{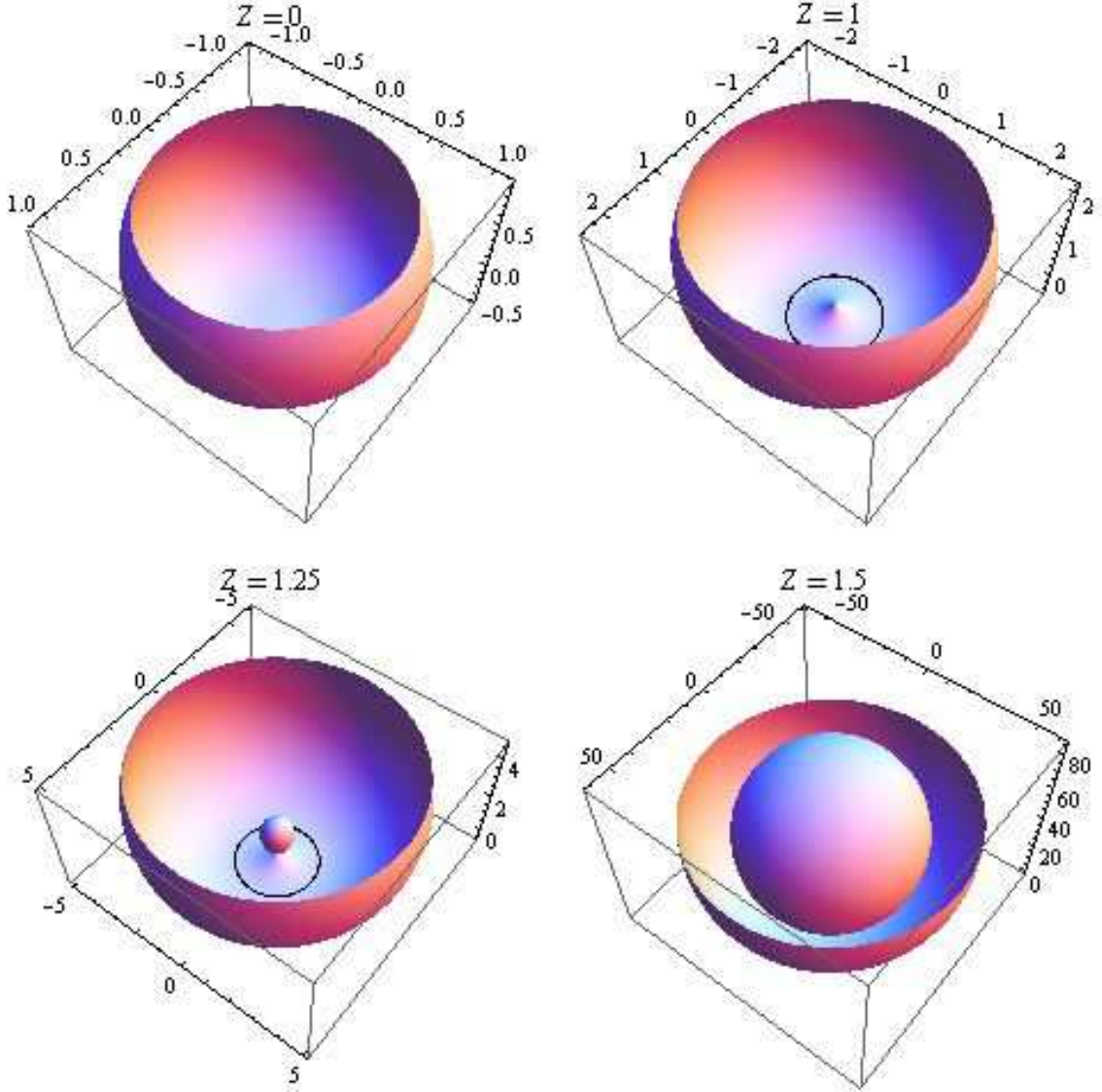}

\end{center}
\caption{Time evolution of spatial geometry for $c=1/2$ (the spatial
region $\psi <\pi /4$ has been chopped off). Each plot shows a fixed-time
projection of the spatial geometry for\ \ $\theta =\pi /2$. Top
left: $z=0$. Top right: $z=1$. Bottom left: $z=1.25$. Bottom right:
$z=1.5$. The black hole event horizon is shown as a black curve.
The top two plots show the spatial geometry before the singularity
has formed. The bottom two plots show the spatial geometry after
the singularity has formed.\ \ }
\label{XRef-FigureCaption-51502747}
\end{figure}

Figure \ref{XRef-FigureCaption-51502747} shows four different snapshots
of the $\theta =\pi /2$ projections of the spatial geometry of the
solution. In the second plot, the spatial geometry is still regular,
but it has an event horizon, shown in black. In the next plot, the
singularity has formed, and a black hole has been created. The protuberance
in the middle of the two plots at the bottom is a spatial region
that lies beyond the singularity, and is shown in the Penrose-Carter
diagram of Figure \ref{XRef-FigureCaption-5307522}\ \ as the spatial
region to the right of the singularity.\ \ \
\begin{figure}[h]
\begin{center}
\includegraphics{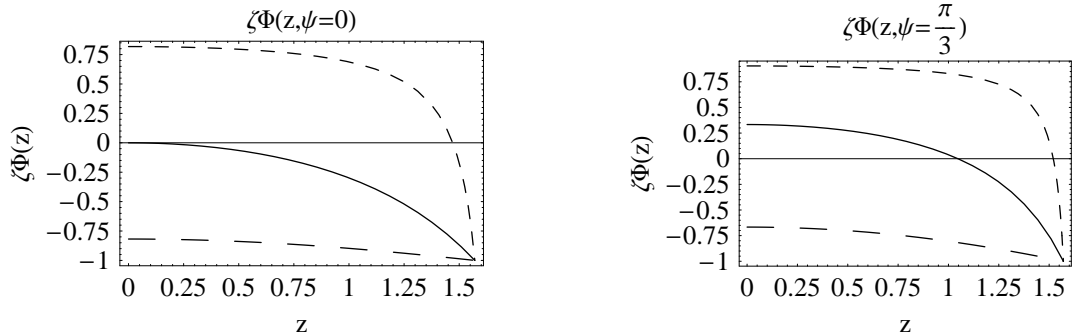}

\end{center}
\caption{Scalar field $\Phi$ as a function of conformal time z for
the two angular positions $\psi$=0 and $\psi$=$\pi$/3. The curves
are for $c=1$(continuous), $c=0.1$ (short dash) and $c=10$(long
dash).}
\label{XRef-FigureCaption-51665528}
\end{figure}
\begin{figure}[h]
\begin{center}
\includegraphics{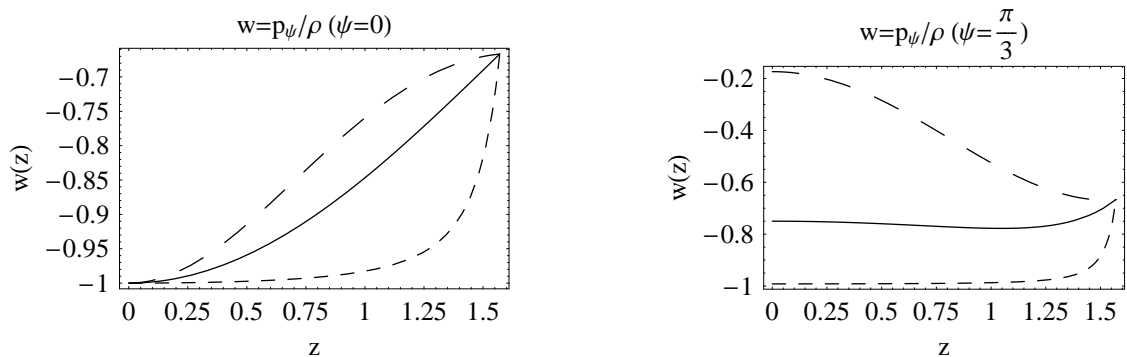}

\end{center}
\caption{Equation of state parameter $w_{\psi }$ as a function of
conformal time $z$ for the two angular positions $\psi$=0 and $\psi$=$\pi$/3.
The curves are for $c=1$(continuous), $c=0.1$ (short dash) and $c=10$(long
dash).}
\label{XRef-FigureCaption-5150107}
\end{figure}

 Figure \ref{XRef-FigureCaption-51665528} plots the scalar field
as a function of conformal time $z$ and angular coordinate $\psi
$ for three different values of the coordinate $c$. Figure \ref{XRef-FigureCaption-5150107}
plots the equation of state parameter $w$ as a function of conformal
time $z$ and angular coordinate $\psi $ for three different values
of the constant $c$.\ \

\section{Conclusions}

In this paper, we generalized two previously known solution generation
techniques for generating minimally coupled Einstein-scalar field
solutions in 4 dimensions (the Buchdahl \cite{Buchdahl:1959nk} and
Fonarev \cite{Fonarev:1994xq} transformations). Two generalizations
were made: i) the transformation was generalized to arbitrary dimension,
and ii) the new transformation allows vacuum solutions with non-zero
cosmological constant as seed. Thus, we are now able to generate
minimally coupled Einstein-scalar field solutions from vacuum solutions
with arbitrary cosmological constant in arbitrary dimension. The
only requirement to a seed solution is that it posesses a hypersurface-orthogonal\ \ Killing
vector field.

We introduced a labeling scheme that labels all solutions in terms
of whether the target solution is invariant with respect to translations
along the Killing vector field (class $\mathcal{S}$\ \ solutions)
or not (class $\mathcal{T}$ solutions). Furthermore, we labeled
the target solution with dimension $N$, as well as whether the seed
solution has non-zero cosmological constant $(\mathcal{S}_{N}^{\Lambda
}, \mathcal{T}_{N}^{\Lambda })$ or not ($\mathcal{S}_{N}^{0}, \mathcal{T}_{N}^{0}$).

The solutions cover new, unknown solutions $(\mathcal{T}_{N}^{\Lambda
})$ as well as previously known solutions, such as the solutions
of Buchdahl $(S_{4}^{0})$, Fonarev ($ \mathcal{T}_{4}^{0}$), Feinstein-Kunze-V\'azquez-Mozo
($\mathcal{T}_{5}^{0}$) and Xanthopoulos \& Zannias \cite{Xanthopoulos:1989kb}
(static, spherically symmetric $S_{N}^{0}$).

The generalization of the Buchdahl transformation to arbtirary dimension
is new $(S_{N}^{0})$. Using the generalized Buchdahl transformation,
we are able to recapture the extra-dimensional static and spherically
symmetric solutions of Xanthopoulos and Zannias as a special case.\ \

 The generalization that allows us to use seed solutions with non-zero
cosmological constant uncovers a new class of Einstein-scalar field
solutions that has previously not been studied. We apply our solution
generation technique in order to study one of the familiy of solutions,
generating Einstein-scalar field solutions from the $(A){\mathrm{dS}}_{4}$
vacuum solutions. The resulting Einstein-scalar field solution that
comes from transforming the de Sitter vacuum is a two-parameter
family of 4 dimensional, inhomogenous, expanding cosmological solutions
with $R\times S^{3}$ topology and exponential scalar potential parametrized
by two parameters: the strength of the scalar potential, $W_{0}$,
and the (positive) cosmological constant $\Lambda $ of the seed
solution. By transforming a solution of this kind to the conformal
frame using the generalized Bekenstein transformation, we find a
spatially finite, expanding and accelerating cosmological solution
that is conformally isometric to the Einstein universe $R\times
S^{3}$. The solution can be parametrized by a cosmologial time scale
$H_{0}$, defined by the strength of the scalar potential, and the
{\itshape inhomogeneity parameter} $c$, defined by the ratio $\Lambda
/W_{0}$. $H_{0}$ defines the scale of the solution, while $c$ defines
the degree of spatial inhomogeneity. Low $c$ implies that the solution
is highly homogeneous. We study null geodesics and find that for
any observer, the solution has a cosmological event horizon at an
angular distance of $\pi /2$ away from the observer. The solution
has an initial singularity, which is a point singularity that vanishes
at early times. The solution is then free of singularities until
a final singularity emerges at late times as a new point singularity.
There are timelike and null geodesics that do not intersect any
of the singularities. For $c<1$, the late time singularity is hidden
inside an event horizon. This family of solutions therefore has
the natural interpretation of being expanding cosmologies in which
a scalar black hole is formed at late times. The energy density
is positive definite for small $c$ ($c<1/8\text{}$) and large $c$
($c>24$). The conformally coupled scalar field satisfies the weak
energy condition as long as the energy density is positive, while
the strong energy condition is generally violated.

\acknowledgments{}

I would like to thank Professor Finn Ravndal at the University of
Oslo for very valuable discussions while preparing this paper.

\appendix

\section{Conformal transformations in arbitrary dimension}
\label{XRef-AppendixSection-98225849}

Here, we will briefly summarize how certain covariant quantities
transform under conformal transformations. For a reference, see
e.g. \cite{Wald:1984-general}. Let $\overline{g}$ and g be the metrics
of two $M$-dimensional space-time geometries that are related by
a conformal transformation $\Omega ^{2}$ as follows:
\begin{gather*}
{\overline{g}}_{\mu \nu } = \Omega ^{2}g_{\mu \nu }
\end{gather*}

Let $f_{\alpha  }$ be an arbitrary covariant $M$-vector. Covariant
derivatives transform as follows under a conformal transformation
$\Omega ^{2}=e^{2\omega }$:
\begin{equation}
\overline{\nabla _{\alpha }}f_{\beta }=\nabla _{\alpha }f_{\beta
}-\left( f_{\beta }\nabla _{\alpha }\omega +f_{\alpha }\nabla _{\beta
}\omega -g_{\alpha \beta }f_{\lambda }\nabla ^{\lambda }\omega \right)
,%
\label{XRef-Equation-21662857}
\end{equation}

where $\overline{\nabla _{\alpha }}$ is the covariant derivative
of the metric $\overline{g}$. Recognizing that $\overline{\nabla
_{\alpha }}U =$ $\partial _{\alpha }U =\nabla _{\alpha }U$, we may
use equation \ref{XRef-Equation-21662857} to compute how second-order
covariant derivatives transform under conformal transformations:
\begin{gather}
\overline{\nabla _{\alpha }} \overline{\nabla _{\beta }}U=\nabla
_{\alpha }\nabla _{\beta }U-\left( \nabla _{\beta }U\nabla _{\alpha
}\omega +\nabla _{\alpha }U\nabla _{\beta }\omega -g_{\alpha \beta
}\nabla _{\lambda }U\nabla ^{\lambda }\omega \right) %
\label{XRef-Equation-2167622}
\\\overline{\Box }U={\overline{g}}^{\alpha \beta
}\overline{\nabla _{\alpha }} \overline{\nabla _{\beta
}}U=e^{-2\omega }( \Box U+\left( M-2\right) \nabla ^{\alpha }U\nabla
_{\alpha }\omega )
\end{gather}

The Ricci tensor $R_{\mu \nu }$transforms as follows under a conformal
transformation $\Omega $:
\begin{equation}
{}{}{\overline{{}R}}_{\mu \nu }\equiv R_{\mu \nu }[ \overline{g}]
= {}{}R_{\mu \nu }[ g] +\frac{2\left( M-2\right) }{\Omega ^{2}}\nabla
_{\mu }\Omega \nabla _{\nu }\Omega -\frac{\left( M-3\right) }{\Omega
^{2}}g_{\mu \nu }\nabla ^{\alpha }\Omega \nabla _{\alpha }\Omega
-\frac{\left( M-2\right) }{\Omega }\nabla _{\mu }\nabla _{\nu }\Omega
-\frac{1}{\Omega }g_{\mu \nu }\Box \Omega %
\label{XRef-Equation-215174554}
\end{equation}

Writing equation \ref{XRef-Equation-215174554} in terms of $\omega$
gives:
\begin{equation}
{}{}{\overline{{}R}}_{\mu \nu }\equiv R_{\mu \nu }[ \overline{g}] =
{}{}R_{\mu \nu }[ g] -\left( M-2\right) \nabla _{\mu }\nabla _{\nu
}\omega +\left( M-2\right) \nabla _{\mu }\omega \nabla _{\nu }\omega
-g_{\mu \nu }( \Box \omega +\left( M-2\right)
\nabla ^{\alpha }\omega \nabla _{\alpha }\omega ) %
\label{XRef-Equation-21664714}
\end{equation}

From equation \ref{XRef-Equation-215174554} we can compute the Ricci
scalar of the transformed metric:
\begin{equation}
\overline{R}\equiv R[ \overline{g}] \equiv {\overline{g}}^{\mu \nu
}{\overline{R}}_{\mu \nu }=\Omega ^{-2}( R[ g] -\left( M-1\right)
\left( \frac{2}{\Omega }\Box \Omega +\left( M-4\right) \nabla
^{\alpha }\Omega \nabla _{\alpha }\Omega \right) )
\end{equation}

It then follows that the Einstein tensor
\[
{\overline{G}}_{\mu \nu }\equiv G_{\mu \nu }[ \overline{g}] =R_{\mu
\nu }[ \overline{g}] -\frac{1}{2}{\overline{g}}_{\mu \nu }R[ \overline{g}]
\]

transforms as follows:
\begin{equation}
{\overline{G}}_{\mu \nu }=G_{\mu \nu }+\frac{3\left( M-2\right)
}{\Omega ^{2}}\left( \Theta _{\mu \nu }[ g,\Omega ,\sqrt{\frac{1}{6}}]
+\frac{\left( M-4\right) }{6}g_{\mu \nu }\nabla ^{\alpha }\Omega
\nabla _{\alpha }\Omega \right) ,%
\label{XRef-Equation-21610338}
\end{equation}

where
\begin{gather}
S_{\mu \nu }[ g,\phi ] \equiv \nabla _{\mu }\phi \nabla _{\nu }\phi
-\frac{1}{2}g_{\mu \nu }\nabla ^{\alpha }\phi \nabla _{\alpha }\phi
\\\Theta _{\mu \nu }[ g,\phi ,\chi ] \equiv S_{\mu \nu }[ g,\phi
] -\chi ^{2}\nabla _{\mu }\nabla _{\nu }\phi ^{2}+\chi ^{2}g_{\mu
\nu }\Box \phi ^{2}.
\end{gather}

Here, g denotes an arbitrary metric,\ \ $\phi $ denotes an arbitrary
scalar function, and $\chi$ is an arbitrary constant.

\section{Generalized Bekenstein transformation}
\label{XRef-AppendixSection-510235039}

The non-minimally coupled Einstein-scalar field action in dimension
$N$ is
\begin{equation}
S[ g,\Phi ,\xi ,\Lambda ,V] \equiv \int d^{N}x\sqrt{-g}\left( \frac{1}{2\kappa
}\left( R-2\Lambda \right) -\frac{1}{2}\nabla ^{\alpha }\Phi \nabla
_{\alpha }\Phi  -\frac{1}{2}\xi ^{2}R \Phi ^{2}-V[ \Phi ] \right)
,
\end{equation}

where $\xi $ is the dimensionless scalar-gravity coupling constant.
We have conformal coupling when $\xi ^{2}\equiv \frac{(N-2)}{4(N-1)}$.
Define $\zeta ^{2}\equiv \kappa  \xi $. When extremalizing this
action, we obtain the conformally coupled Einstein-scalar field
equations:
\begin{gather}
G_{\mu \nu }[ g] =\frac{\kappa }{\left( 1-\zeta ^{2}\Phi ^{2}\right)
}\left( \Theta _{\mu \nu }[ g,\Phi ,\xi ]  - g_{\mu \nu }W[ \Phi
] \right) %
\label{XRef-Equation-510233227}
\\\Box \Phi -\xi ^{2}R \Phi -\frac{d W[ \Phi
] }{d \Phi }=0,%
\label{XRef-Equation-510233249}
\end{gather}

where $R$ is the Ricci scalar of the solution geometry and we have
introduced the effective scalar potential $W[ \Phi ] \equiv V[ \Phi
] +\kappa ^{-1}\Lambda $.

Let $(\overline{g},\phi ,\overline{W})$ be an $N$ dimensional solution
to the minimally coupled Einstein-scalar field equations, and let
$(g, \Phi , W)$ be an $N$ dimensional solution to the non-minimally
coupled Einstein-scalar field equations with arbitrary scalar-gravity
coupling $\xi $. A {\itshape Bekenstein transformation} $(\Omega
, f)$ is defined as a conformal transformation $\Omega ^{2}$ mapping
the metric $g$ into $ \overline{g}$\ \ and a function $f$ mapping
the scalar field $\Phi $ into $\phi $:
\begin{gather*}
{\overline{g}}_{\mu \nu }=\Omega ^{2}g_{\mu \nu }
\\\phi =f[ \Phi ]
\end{gather*}

Now, let us state, without proof, the generalized Bekenstein theorem
in $N$ dimensions:
\begin{theorem}

{\itshape For every minimally coupled Einstein-scalar field solution,
there are two, and only two, Bekenstein transformations that relate
the minimally coupled solution to two Einstein-conformal scalar
field solutions. If }$(\overline{g},\phi ,\overline{W})${\itshape
is an }$N${\itshape  dimensional solution to the Einstein-scalar
field equations with a minimally coupled scalar field }$\phi ${\itshape
and a scalar field potential }$\overline{W}$, {\itshape there are
two independent solutions }$(g, \Phi , W)$, labeled A and B, {\itshape
to the Einstein-conformal scalar field equations, eqs.} \ref{XRef-Equation-510233227}{\itshape
and} \ref{XRef-Equation-510233249}{\itshape ,\ \ given by}\label{XRef-Theorem-510234826}
\begin{gather}
g_{\mu \nu }=\Omega ^{-2}{\overline{g}}_{\mu \nu }
\\\Phi =\begin{cases}
\frac{1}{\zeta }\tanh  \zeta  \phi  & \left( A\right)  \\
\frac{1}{\zeta }\coth  \zeta  \phi  & \left( B\right)  \\
\end{cases},
\\\Omega =|1-\zeta ^{2}\Phi ^{2}|^{\frac{1}{N-2}}=\begin{cases}
{\left( \cosh  \zeta  \phi \right) }^{-\frac{2}{N-2}} & \left( A\right)
\\
{\left( \sinh  \zeta  \phi \right) }^{-\frac{2}{N-2}} & \left( B\right)
\\
\end{cases}
\\W[ \Phi ] =\begin{cases}
\Omega ^{N}\overline{W}[ \phi [ \Phi ] ] =|1-\zeta ^{2}\Phi ^{2}|^{\frac{N}{N-2}}\overline{W}[
\phi [ \Phi ] ]  & \left( A\right)  \\
-\Omega ^{N}\overline{W}[ \phi [ \Phi ] ] =-|1-\zeta ^{2}\Phi ^{2}|^{\frac{N}{N-2}}\overline{W}[
\phi [ \Phi ] ]  & \left( B\right)  \\
\end{cases}
\end{gather}
\end{theorem}\label{Def-seed}\label{ClassT00Form}

\end{document}